\documentclass{aa}
\usepackage{txfonts}
\usepackage{fullpage,graphicx}
\usepackage{setspace}
\usepackage{subfigure}
\usepackage{natbib}

\begin{document}

\title{Was a cloud-cloud collision the trigger of the recent star formation in Serpens?} 


\author{A. Duarte-Cabral\inst{1}\thanks{Funded by the Funda{\c{c}}{\~a}o 
  para a Ci{\^e}ncia e a Tecnologia (Portugal)}  
  \and C. L. Dobbs\inst{2,3,4}  
  \and N. Peretto\inst{1,5}  
  \and G. A. Fuller\inst{1}}

\offprints{Ana Duarte Cabral, \email{Ana.Cabral@postgrad.manchester.ac.uk}}

\institute{Jodrell Bank Centre for Astrophysics, School of Physics and
  Astronomy, University of Manchester, Oxford Road, Manchester, M13 9PL, U.K.
  \and School of Physics, University of Exeter, Exeter EX4 4QL, U.K.
  \and Max-Planck-Institut f\"ur extraterrestrische Physik, Giessenbachstra\ss{}e, D-85748 Garching, Germany
	\and Universit\"ats-Sternwarte M\"unchen, Scheinerstra\ss{}e 1, D-81679 M\"unchen, Germany
  \and Laboratoire AIM, CEA/DSM-CNRS-Universit\'e Paris Diderot, IRFU/Service d'Astrophysique, 
  C.E. Saclay, Orme de merisiers, 91191 Gif-sur-Yvette, France}
  
\date{Received 27 July 2010 / Accepted 10 January 2011}

\abstract {The complexity of the ISM is such that it is unlikely that star formation is initiated in the same way in
  all molecular clouds. While some clouds seem to collapse on their own, others may be triggered by an
  external event such as a cloud/flow collision forming a gravitationally unstable enhanced density
  layer.}  {This work tests cloud-cloud collisions as the
  triggering mechanism for star formation in the Serpens Main Cluster as has
  been suggested by previous work.}  {A set of smoothed particle hydrodynamics (SPH)
  simulations of the collision between two cylindrical clouds are performed and compared to 
  (sub)millimetre observations of the Serpens Main Cluster.} 
  {A configuration has been found which reproduces {many of the
  observed characteristics of Serpens, including some of the main features of the peculiar velocity field}. The
  evolution of the velocity with position throughout the model is similar
  to that observed and the column density and masses within the modeled 
  cloud agree with those measured for the SE sub-cluster. {Furthermore, our results also show that 
  an asymmetric collision provides the ingredients to reproduce lower density filaments 
  perpendicular to the main structure, similar to those observed}. {In this scenario,} the formation
  of the NW sub-cluster of Serpens can be reproduced only if there is a
  pre-existing marginally gravitationally unstable region at the time the
  collision occurs.}  {This work supports the interpretation that a collision
  between two clouds {may have been the trigger of} the most recent burst of star formation in
  Serpens. It not only explains the complicated velocity structure seen in the
  region, but also the temperature differences between the north (in
  ``isolated'' collapse) and the south (resulting from the shock between the
  clouds). In addition it provides an explanation for the sources in the south
  having a larger spread in age than those in the north.}

\maketitle 


\section{Introduction}
\label{intro}
\label{fmc}

{Observations of molecular clouds indicate that stars form in dense, clumpy filaments \citep[e.g.][]{2010A&A...518L.102A,2010A&A...518L.100M}. However, the processes controlling the formation of these structures and their role in triggering specific star formation episodes are widely debated. One process could simply be gravity controlled: quasi-static or dynamic gravitational collapse of non-spherical clouds can easily generate filamentary structures which subsequently fragment to form stars \citep[e.g.][]{2002ApJ...581.1194P,2007ApJ...654..988H}. {The contraction of an idealised isolated turbulent cloud leads to the formation of small scale filaments ($\sim$~0.1~pc long) within which stars begin to form, such as shown by simulations from \citet[][and \citeyear{2009MNRAS.397..232B}]{2003MNRAS.339..577B,2005ApJ...620..786K,2009MNRAS.392..590B}.}
 Alternatively, external triggers can compress molecular gas to generate high-density fragmented filaments. Such triggers can be of very different natures: ionisation/shock fronts around OB stars \citep[e.g.][]{1977ApJ...214..725E,2007MNRAS.375.1291D}, large scale colliding flows generated by supernovae/galactic shear \citep[e.g.][]{2008ApJ...674..316H}, or even molecular cloud collisions \citep[e.g.][, \citeyear{2009A&A...504..437A}]{2003MNRAS.340..841G,2007MNRAS.378..507K,2009A&A...504..451A}.}

{Each of these processes have been tested using numerical simulations, and their respective efficiency in forming stars have been investigated. For some local clouds, we are beginning to acquire sufficiently detailed observations to generate a more informed picture of how star formation has progressed in those clouds. With these observations it becomes possible to  perform dedicated simulations to model and better understand the processes at work. Only few such studies have been performed so far. For instance, \citet{1999ApJ...527..285B} used the results of their  turbulent colliding flow simulations to argue that Taurus-Auriga formed via such a process; \citet{2007A&A...464..983P} used hydrodynamical simulations to model the global collapse and fragmentation of the NGC~2264-C protocluster; and \citet{2009ApJ...704.1735H} compared  flow-driven models for the formation of isolated molecular clouds with the Pipe nebula. In this paper, we use numerical simulations to investigate whether the recent star formation episode observed in the Serpens Main Cluster could have been triggered by the collision of two molecular clouds, as proposed in \citet[][]{2010A&A...519A..27D}. }

Serpens is an interesting region in several aspects. It is a filamentary region with two
compact young star forming clumps. The gas emission reveals an interesting velocity structure and {has an unusual temperature structure} \citep[][]{2010A&A...519A..27D}. The observations point to remnant signatures of the trigger behind the current young on-going star formation of Serpens. \citet[][]{2010A&A...519A..27D} suggested that this region appears to have been affected by a collision of two clouds or flows. Such a scenario is also suggested by recent studies of the global dynamics of Serpens and other star forming regions \citep[e.g.][]{2010ApJ...719.1813H,2010A&A...520A..49S,2010arXiv1004.2466G}. 


Here we use smoothed particle hydrodynamics (SPH) simulations of cloud-cloud
collisions to attempt to reproduce the column density distribution and dynamical
properties of the Serpens Main Cluster. 
A brief summary of the observationally inferred characteristics of Serpens and
the motivation for the simulations are presented in Section~\ref{data}. Section~\ref{simul}
explains the SPH code, the initial conditions and the geometrical
configurations adopted. The results of the calculations are presented in
Section~\ref{res}. Section~\ref{discussion} presents a discussion of these results, where we compare
the numerical results with the observations and choose the \textit{best-fit}
scenario to represent the region. The final remarks and conclusion
are laid out in Section~\ref{conc}. {Finally, in Appendix~\ref{snapshots} (available online only) we present temporal snapshots for all the models presented in this work.}


\section{The structure of the Serpens Main Cluster region}
\label{obser}
\label{data}

\begin{figure}[!t]
\centering
\includegraphics[angle=270,width=0.4\textwidth]{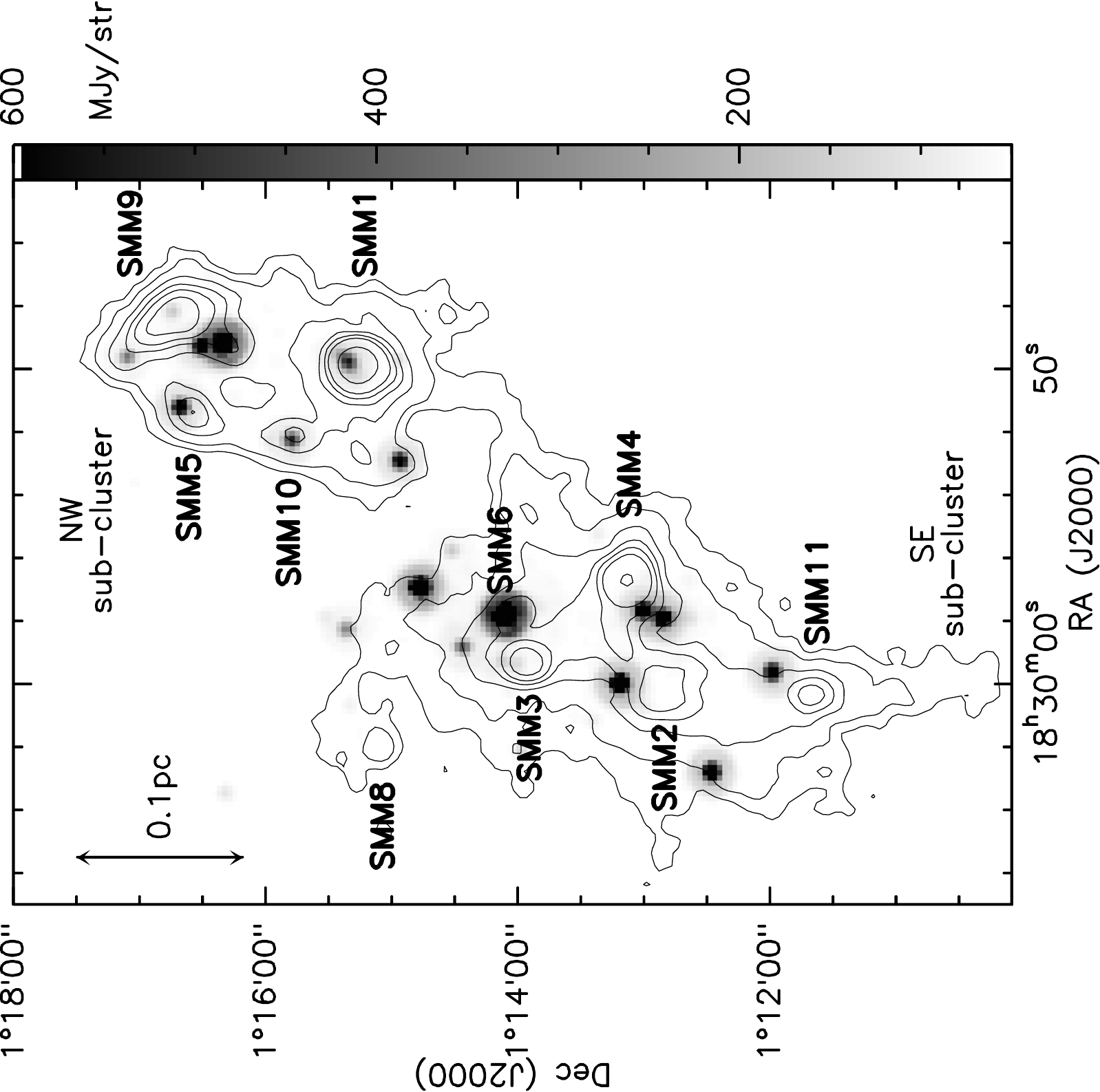}
\caption{\small{Serpens region as seen in 850~$\mu$m dust continuum emission with SCUBA \citep[][]{1999MNRAS.309..141D}
	in contours, at 0.4, 0.6, 1, 1.4, 1.8, 2.4 and 5~Jy~beam$^{-1}$. In grey scale is the 24~$\mu$m from Spitzer MIPS
  \citep[][]{2007ApJ...663.1139H}. All the sources seen in this image, both 24~$\mu$m and 
  850~$\mu$m sources, are Class 0 and I, with only a few flat-spectrum sources.
  The 850~$\mu$m sources are labeled.}}
\label{fig:850micron}
\end{figure}

Serpens Main Cluster is a low mass star forming region at 260~pc from the Sun
\citep[][]{1996BaltA...5..125S}. It comprises two compact protoclusters,
hereafter referred to as sub-clusters, lying in a 0.6~pc long filamentary structure, along a NW - SE direction.
At first sight, the NW and SE sub-clusters appear rather similar. The dust
emission \citep[][ and Fig. \ref{fig:850micron}]{1999MNRAS.309..141D}, shows
that they have similar masses within similar sized regions: $\sim$
30~M$_{\odot}$ in 0.025~pc$^{2}$ each, as estimated from both gas emission
from C$^{18}$O \citep[][]{2010A&A...519A..27D} and 850~$\mu$m continuum emission
from SCUBA. Furthermore, they both have sources at roughly the
same stage of evolution, Class 0 and Class I protostars, powering a number of
outflows \citep[][]{1999MNRAS.309..141D,2010MNRAS...Graves}.

Previous studies \citep[][]{2004A&A...421..623K,2006ApJ...644..307H} had
estimated an average age of 10$^{5}$~yr for these sources, possibly marking a
triggering event. Older sources are also found in the field, but they appear
more dispersed, with no obvious connection to the current protostars, and no
longer embedded nor surrounded by any cold dust seen at submillimetre
wavelengths. It is thought these belong to a different burst of star formation
which occurred 2~Myr before the current one, and that these sources are now in
the field as an unbound cluster.

\begin{figure}[!t]
\centering
\includegraphics[angle=270,width=0.4\textwidth]{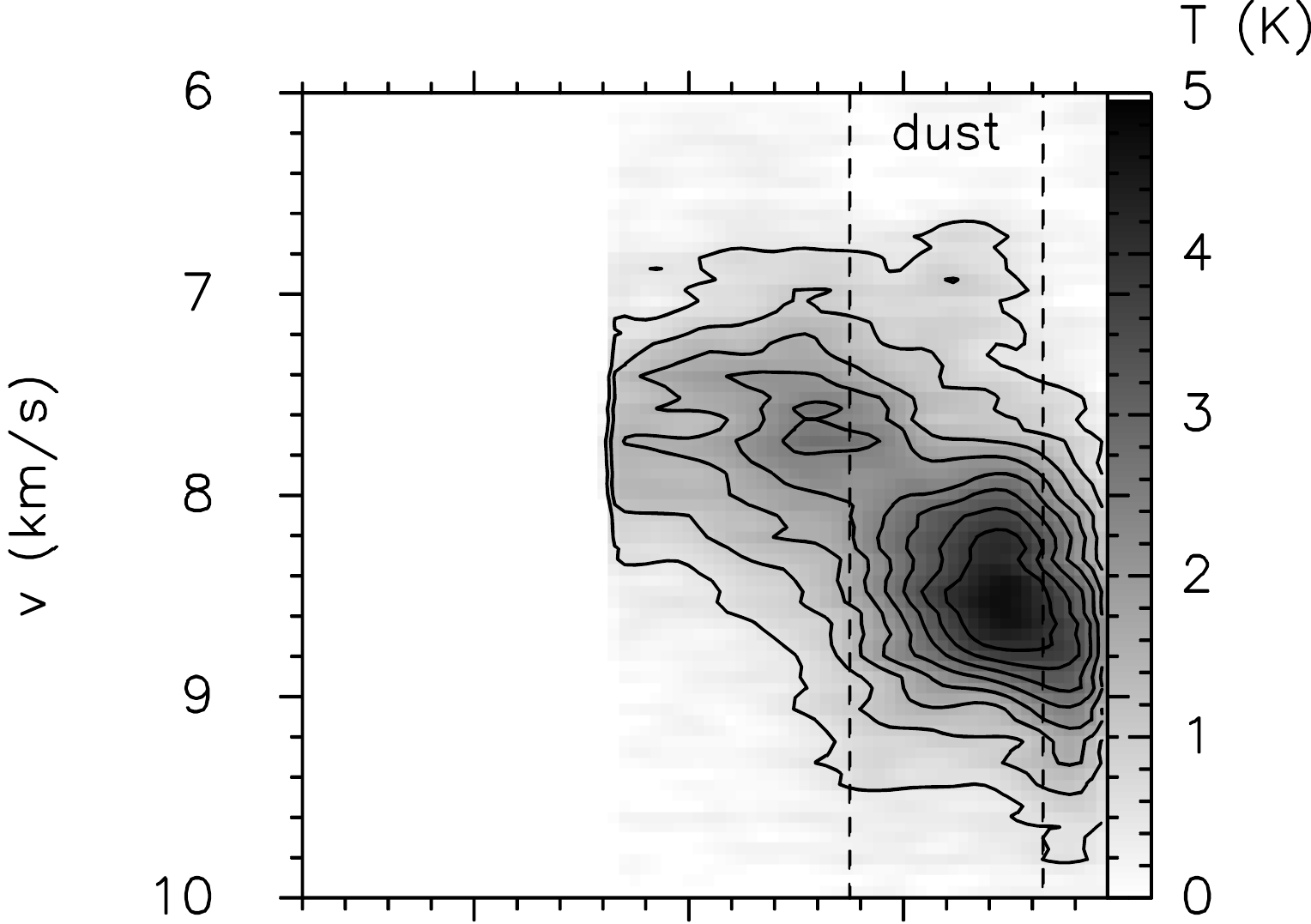}\\
\includegraphics[angle=270,width=0.4\textwidth]{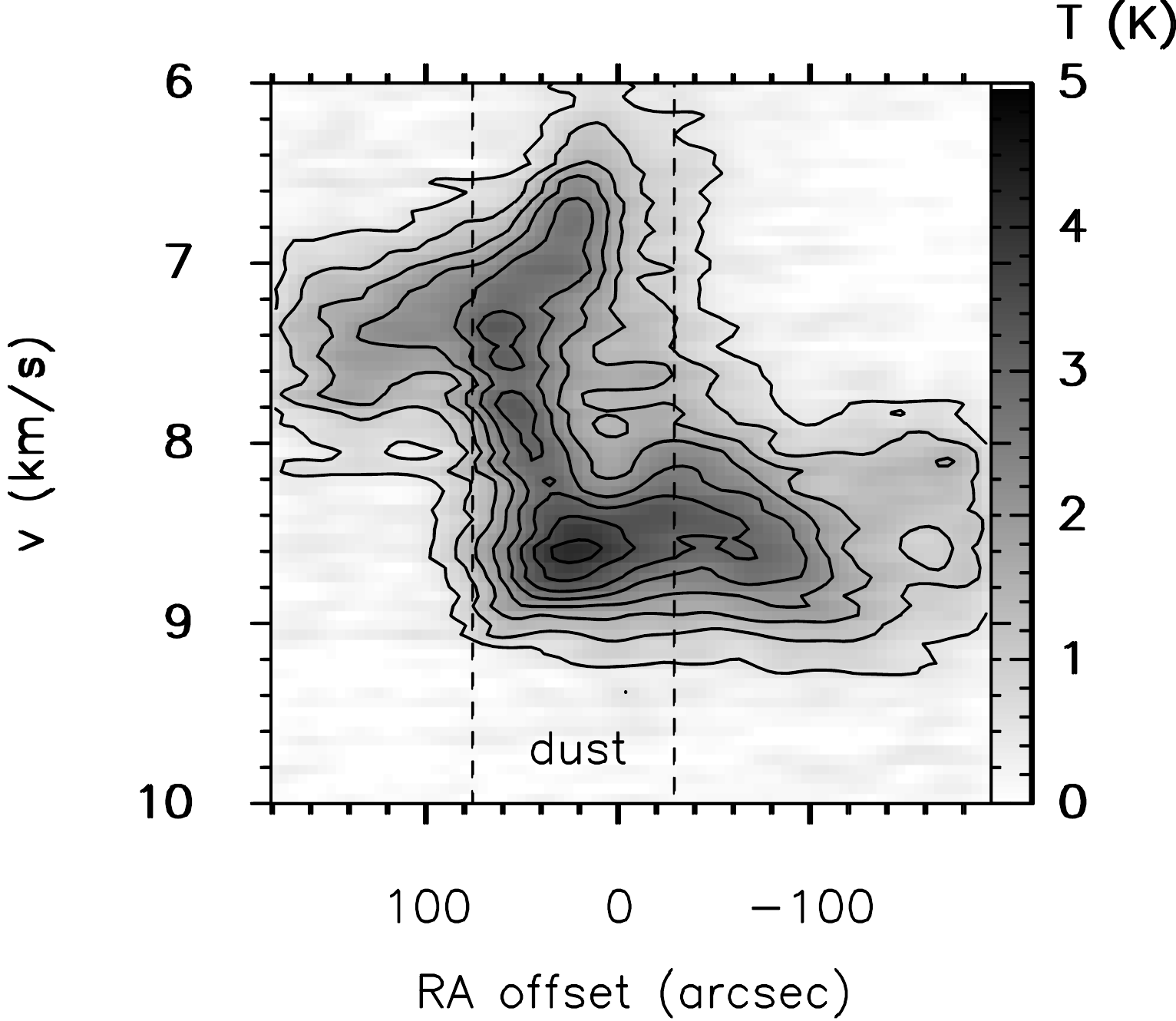}
\caption{\small{Position-velocity diagrams at constant declination of C$^{18}$O J=1$\rightarrow$0 
	(grey scale and contours) as examples of the {typical velocity structure in the NW sub-cluster (top) and the SE sub-cluster (bottom)} \citep[][]{2010A&A...519A..27D}. Right Ascension varies from $18^{h}$$30^{m}$$06^{s}$ to $18^{h}$$29^{m}$$46^{s}$ for all panels. Declinations are $1^{\circ}$$16'$$48''$ (top panel) 
and  $1^{\circ}$$12'$$38''$ (lower panel). {The positions of these cuts in the map are shown as white lines in Fig.~\ref{fig:bluered}.} The dashed lanes labeled as ``dust'' represent the regions whose 850$\mu$m emission is above 0.6~Jy~beam$^{-1}$}.}
\label{fig:pvobs}
\label{fig:pvobs2}
\end{figure}

\begin{figure}[!t]
\centering
\includegraphics[width=0.35\textwidth]{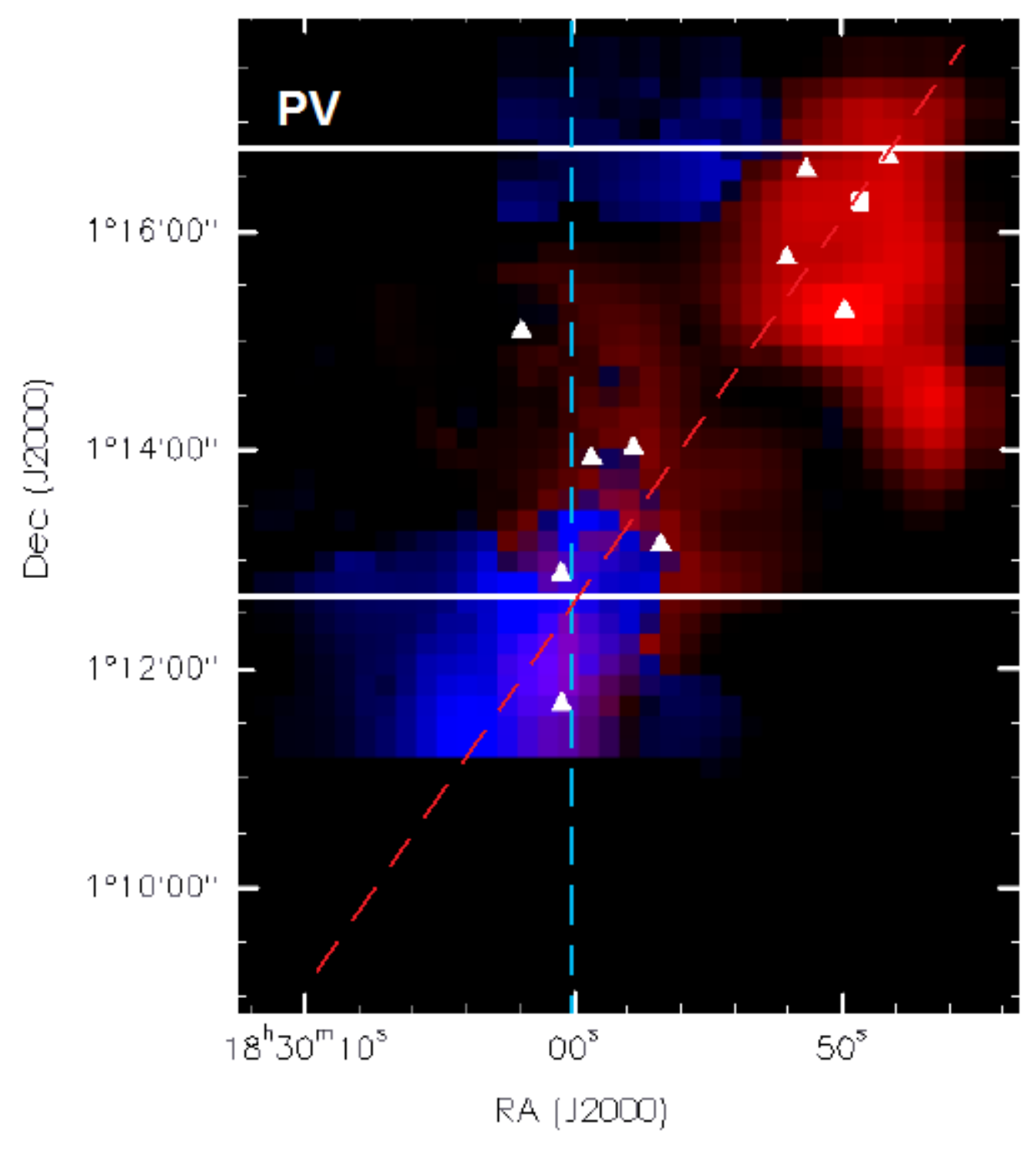}
\caption{\small{C$^{18}$O J=1-0 emission separated into blue and red components,
  using the line fitting and splitting by \citet[][]{2010A&A...519A..27D}. The
  blue emission corresponds to the total integrated intensity of the low velocity
  cloud and the red emission corresponds to the high velocity cloud. 
  The submillimetre sources from the 850~$\mu$m emission \citep[][]{1999MNRAS.309..141D} are marked with white
  triangles and serve as a guide to the location of the dust emission. {The position 
  of the PV diagrams in Fig.~\ref{fig:pvobs} are shown by the white solid lines, and the main axis of the blue shifted and red shifted cloud is shown as blue and red dashed line respectively.}}}
\label{fig:bluered}
\end{figure}

In contrast to the similarities of the dust emission, 
the velocity structure and molecular emission from each sub-cluster
are strikingly different. 
The NW is a well ``behaved'' sub-cluster with one main velocity throughout, 
matching the systemic velocity of the cloud with only a smooth velocity shift towards the
edges of the sub-cluster. This is shown in the C$^{18}$O
J=1$\rightarrow$0 position-velocity (PV) diagram of Fig.~\ref{fig:pvobs} (top) and on Fig.~\ref{fig:bluered}
which shows the structure of the higher and lower velocity components of Serpens. 
In this NW sub-cluster, the gas and dust emission have similar distributions, with the gas emission peaks
corresponding to dust emission peaks. In addition, Spitzer 24~$\mu$m sources
are coincident with many of the 850~$\mu$m sources, hereafter submillimetre sources
(Fig.~\ref{fig:850micron}).

In contrast, the SE sub-cluster is quite different and its velocity structure is more 
complex. {Optically thin tracers show locations with a broad single component (at the north of the SE sub-cluster) 
which then splits into two clearly separated components (Fig.~\ref{fig:bluered} and lower panel of Fig.~\ref{fig:pvobs2}). At the southern end of the sub-cluster, the filamentary structure is then only seen in one velocity component along the line of sight  \citep[lower PV diagram of Fig.~17 of][]{2010MNRAS...Graves}.}
Furthermore, the gas and submillimetre emission peaks do not coincide. Rather
the gas emission appears to peak between the submillimetre sources
\citep[][]{2010A&A...519A..27D}. Finally, there are only a few 24~$\mu$m sources
associated with the submillimetre dust continuum emission
(Fig.~\ref{fig:850micron}). The overall picture resembles that of a region
where star formation is an on-going process, with some younger sources (the
purely submillimetre sources) and others older (the purely 24~$\mu$m
sources), unlike the NW sub-cluster where the sources appear to be all at
the same evolutionary stage.

A final difference between the two Serpens sub-clusters is the gas temperature
\citep[][]{2010A&A...519A..27D}. The NW has a very uniform temperature around
10K, while the SE has both higher and more varied temperatures, ranging
between 10 and 20K. In addition, the temperature peaks are not coincident with
any submillimetre source but rather between them, in the region where the two
velocity components overlap.

\label{motivation}

Thus the differences between the two sub-clusters indicate that they have had 
a complicated history. A scenario capable of explaining the trigger
of Serpens star formation has to reproduce the inhomogeneities in the sources' age, 
velocity structure and temperature distribution.
The collision of two filament-like clouds, colliding only over a
portion of their length could provide such a trigger.


\section{SPH Simulations}
\label{simul}

To test the proposal that the structure and star formation in the Serpens Main
Cluster results from triggering by a cloud-cloud collision we have performed a
set of SPH simulations which are compared in detail 
with millimetre and submillimetre observations of Serpens.

\label{compare}

\subsection{Numerical Code}
		
The calculations in this paper are performed using a SPH code based on a version by 
\citet[][]{1990ApJ...348..647B}, but has been subject to substantial 
modifications, including sink particles \citep[][]{1995PhDT.......181B}, 
variable smoothing lengths \citep[][]{2005MNRAS.364..384P} and magnetic fields 
\citep[][]{2007MNRAS.377...77P}. The code has been frequently used for 
simulations of star formation \citep[e.g.][]{2006MNRAS.371.1663D,2009MNRAS.392..590B}.

We perform calculations which include the hydrodynamics and self gravity, but do
not include magnetic fields. We use sink particles, which are inserted in
regions of high density ($10^{-12}$ g cm$^{-3}$) that are undergoing collapse,
to represent protostars. However for all the analysis in this paper we use the
frame of the simulation when the first sink particle appears, i.e. before
stellar feedback is likely to have an effect. Therefore our results are not
dependent on the details, or dynamics of the sink particles.

For all the calculations we adopt an isothermal equation of state. Again, we
are only interested in the evolution of the clouds until star formation takes
place. So this simplification is reasonable given that we consider the
structure of the gas prior to stellar feedback.


\subsection{Initial conditions}
\label{initialc}

{Given the elongated appearance of both the high and low velocity clouds in Serpens, we hypothesised the velocity structure of the Serpens cluster to be due to the collision of two elongated clouds {($\sim$~1~pc long)}. Cloud collisions have long been thought to be an important process in the ISM \citep[since][]{1954BAN....12..177O}. They are potential triggers for star formation \citep[e.g.][]{1986ApJ...310L..77S,1995AJ....110.2256V}, and are frequently found in galactic scale simulations \citep[e.g.][]{2008MNRAS.391..844D,2009ApJ...700..358T}. {We do not suggest that cloud collisions are responsible for all star formation in the Galaxy, but aim to test how well this scenario can model the specific case of Serpens.} Taking individual clouds from galactic simulations to model the collision is beyond the scope of this paper, and furthermore does not allow any freedom of the geometry of the clouds or collision. Instead, for our models we assume a simplified initial configuration with cylindrical clouds. The choice of cylinders is reasonable given that the most commonly observed aspect ratio of molecular clouds is $\sim$~2 \citep[][]{2006ApJ...638..191K}. Thus our model is a simplified but plausible scenario in the local Galaxy.} 

For our calculations, we based the properties of the cylinders on the observations of Serpens, but with the requirement that the cylinders are not too far from virial equilibrium. We required the star formation to be primarily driven by the collision, and preferably the cylinders should retain their elongated shape as much as possible prior to the collision. Based on the observed spatial distribution of the Serpens' high and low velocity clouds, all the simulations we run involve two cylindrical clouds with radii of 0.25~pc. {For all the calculations we have kept one cylinder vertical and the other tilted and, with the exception of run D (Table~\ref{tab:configs}), the cylinders have a length of 0.75 and 1~pc respectively.} 

To ensure the stability of the cylinders prior to the collision, we used the formula for stability of finite cylinders given in \citet[][]{1983A&A...119..109B} to estimate the masses of the cylinders,
\begin{equation}
J=\frac{GM f(L/D)}{L R_gT/\mu}
\label{eq:stability}
\end{equation}
where $R_g$ is the gas constant, $G$ is the gravitational constant, $M$ is the mass of the cylinder, $L$ the length, $T$ the temperature and $\mu$ the mean molecular weight (we assumed $\mu=2.0$ for these calculations). $f(L/D)$ is a function measuring the shape of the cloud, but is of order unity for the dimensions of our cylinders. The cylinder is stable providing $J~\lesssim~0.8$, a condition which reduces to approximately $M(\rm{M_{\odot}})/T(\rm{K})~\lesssim~0.8$ for a 1 pc cylinder.
In addition, we {include an external pressure term, which is subtracted from each SPH particle when calculating pressure gradients \citep[][]{2009MNRAS.398.1537D}. This represents a confining pressure from a low density medium}, which minimises diffusion of the outer parts of the cylinders, particularly in the runs including turbulence.

To mimic the supra-thermal line widths observed in Serpens each model was also run for the same initial configuration adopting a turbulent velocity field for the particles. The turbulent field was set up using the method of \citet{2002MNRAS.332L..65B}, and described in depth in \citet{1995ApJ...448..226D}. The energy of the turbulent field $E(k)$ is chosen to follow a power law of $E(k) \propto k^{-4}$, with $k$ being the wavenumber. This gives, for a given scale, $r$, a velocity dispersion, $\sigma_{v}$, such that $\sigma_{v} \propto r^{0.5}$. This is similar to the observed Larson relations for clouds \citep[][]{1981MNRAS.194..809L}, although we do not aim to capture the precise details of the internal motion, but rather to compare results with and without turbulence.

\begin{figure*}[!t]
\centering
\includegraphics[width=\textwidth]{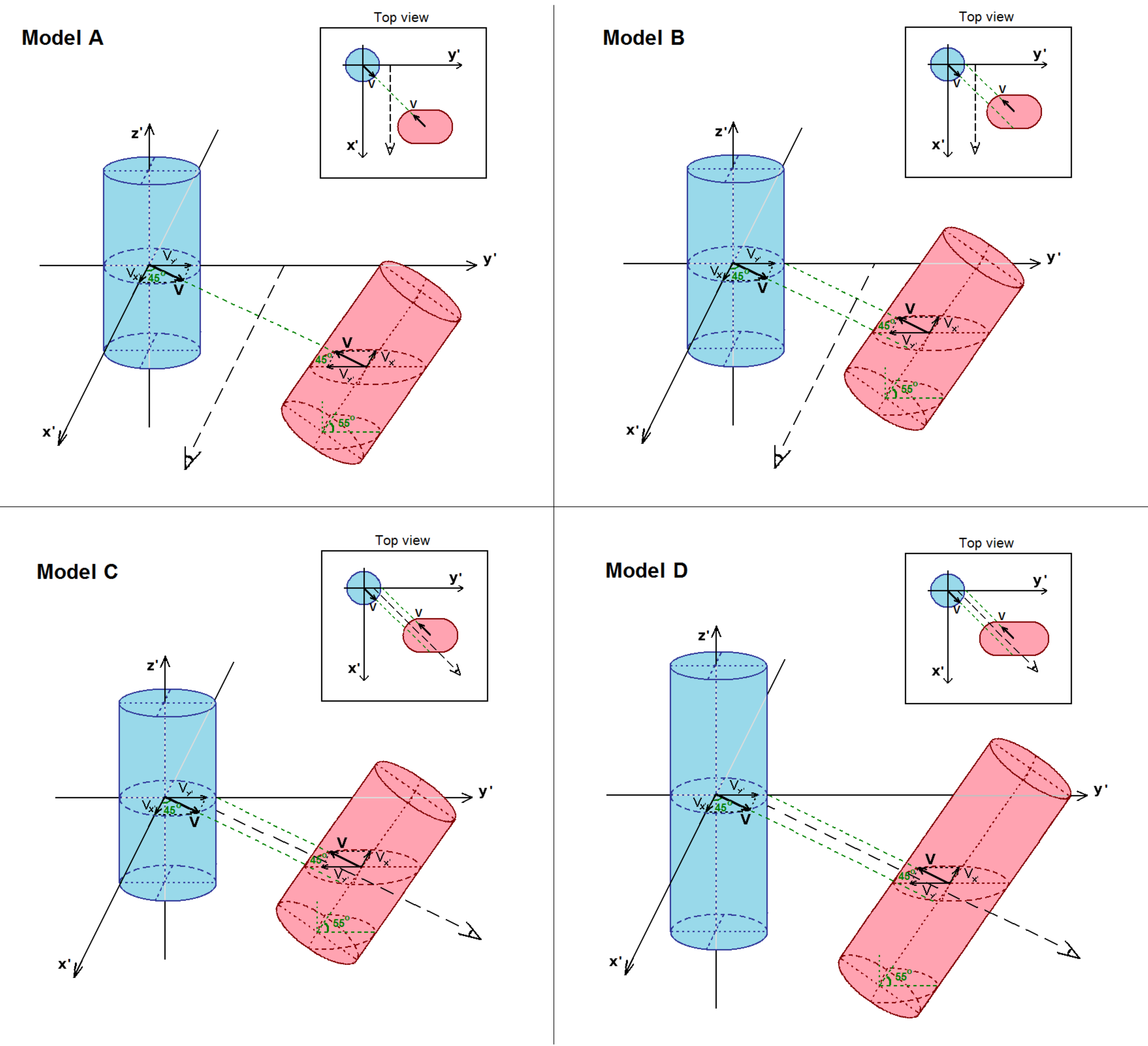}
\caption{\small{{Illustrative diagrams showing the starting configuration of the cylinders for the direct collision (model A, top left panel), offset collision (model B, top right panel; and model C, lower left panel) and offset collision with longer cylinders (model D, lower right panel). The black dashed line represents the line of sight, and the green dashed lines are the direction of the motion of the cylinders. On the top right corner of each panel there is a sketch with the configuration as viewed from the top.}}}
\label{fig:cylA}
\label{fig:cylB}
\label{fig:cylC}
\label{fig:cylD}
\end{figure*}

From the velocity-space distribution of C$^{18}$O emission in the Serpens Main
Cluster (see Fig.~\ref{fig:bluered} and \citealt[][]{2010A&A...519A..27D}), we find
that the higher velocity component (red in Fig.~\ref{fig:bluered}) is tilted
at 55$^{\circ}$ angle from the horizontal plane, and is mostly to the East on the
map. The lower velocity component (blue in Fig.~\ref{fig:bluered}) is roughly
vertical and on the West of the map (overlapping only along the SE end of the
filament of star formation). {To match the observations, the higher velocity component 
corresponds to the longer and tilted cylinder moving away from us on the right-hand side (see Fig.~\ref{fig:cylA} red cylinder). This cylinder has an elevation angle of 55$^{\circ}$ above the horizontal plane and an azimuth angle of 
0$^{\circ}$, i.e. the axis is in a plane parallel to the $y'z'$ plane. The lower velocity component 
corresponds to the vertical cylinder which is coming toward us from the left-hand side 
(see Fig.~\ref{fig:cylA} blue cylinder).}

The velocities of the cylinders are chosen to agree with the velocity difference of the gas in the Serpens cluster (see lower panel on Fig.~\ref{fig:pvobs2}), which would correspond to the velocities of the clouds during the collision. {The initial velocities of each cylinder are $v=\pm 1$~kms$^{-1}$, i.e. $v_{x^\prime}$ and $v_{y'}=\pm$0.77~kms$^{-1}$}. The cylinders have no initial net velocity in the $z$ direction (and without turbulence the particles have zero 
velocity in the $z$ direction).


 
Each of the calculations were run with and without a turbulent velocity field. The differences between the four runs presented in this paper are sketched in Figure~\ref{fig:cylA}, and summarised in Table~\ref{tab:configs} where ``Collision'' specifies if the collision is off-centered or head-on with respect to the centre of the cylinders; ``LOS $\sphericalangle$ V'' represents the angle between the line of sight (LOS) and the direction of the cylinders' motion; ``Length'' is the height of the longer (i.e. the tilted) cylinder in each run; ``Velocities'' specify the initial distribution of velocities: ``Non-Turb'', means zero velocity dispersion in the initial conditions and ``Turb'' indicates a turbulent-generated velocity dispersion amplitude of 0.5~kms$^{-1}$ (in each direction). This is similar to the observed turbulent velocity dispersion in Serpens of the order of 0.5~kms$^{-1} $\citep[][]{2010A&A...519A..27D}

All other properties of the cylinders, such as the temperatures and densities, are fixed. In the ``short-cylinder" runs (A, B and C) we used 500,000 particles in total, with 250,000 particles in each cylinder. {Initially, the particles are uniformly randomly distributed, 
to provide a roughly uniform density distribution within each cylinder.}

As discussed in \citet[][]{2010A&A...519A..27D}, the NW sub-cluster tends to have a uniform temperature of 10~-~13~K. On the other hand the temperature of the SE sub-cluster is more difficult to estimate, given that there are two clumps along the line of sight. \citet[][]{2010A&A...519A..27D} obtained gas temperatures between 10 and 20~K. {However, as shown by Eq.~\ref{eq:stability}, a kinetic temperature of 30~K is required for a 30~M$_{\odot}$ cylinder to be relatively stable. Hence we decided to add in the simulations this extra internal pressure in order to avoid the collapse of the cylinders before the collision. We chose masses of 30 and 45 M$_{\odot}$ for the two cylinders, and verified that for a temperature of 30~K the cylinders did not collapse prior to the collision. This does not have any impact on the velocity and column density structure discussed in the rest of the paper.} \\

We set the cylinders to collide at an angle of 45$^{\circ}$ relative to the line of
sight. In addition, we chose to position the cylinders so that either they collide head on (run A), or their major axis are offset by 0.25~pc (run B). In the direct collision, model A, we centre one cylinder at cartesian coordinates (0,0,0)$\:$pc and the second (longer) cylinder at (0.8,0.8,0)$\:$pc ({Fig.~\ref{fig:cylA} top left panel}). For the offset collision, model B, the second cylinder is instead placed at (0.8,1.05,0)$\:$pc. We illustrate the offset configuration in Fig.~\ref{fig:cylB} {(top right panel)}.

The results from the runs A and B are summarised in Sections~\ref{nonT} to
\ref{turb}. These runs, however, show that by the time sink particles are
formed the two cylinders fully overlap along the line of sight, even though 
they did not interact everywhere. This means that we can see two 
velocity components co-existing in most regions
(e.g. Fig.~\ref{fig:pvturb} top left panel), inconsistent with the
observations where the two velocity components overlap only where the
collision takes place.

Therefore, we changed the perspective so that the motion of the
cylinders is purely along the line of sight, model C (Fig.~\ref{fig:cylC} bottom left panel), to
restrict the region where the two-components overlap to where the
cylinders interact directly. {Finally, model D
(Fig.~\ref{fig:cylD} bottom right panel) aims to reproduce the
whole extent of the observed cloud and, therefore, it is similar to model C, except the
cylinders were made one and half times longer by extending both the top and bottom of the cylinders.} The number of
particles was correspondingly increased in this calculation.

\begin{table}[!t]
\caption{Configuration of the different simulations.}
\label{tab:configs}
\begin{tabular}{l|c|c|c|l}
\hline
\hline
Run & Collision & LOS $\sphericalangle$ V & Length & Velocities \\
\hline
A$_{\mathrm{non-T}}$   & Direct & 45$^{\circ}$ & 1pc & Non-Turb \\
A$_{\mathrm{T}}$       & Direct & 45$^{\circ}$ & 1pc & Turb \\
\hline
B$_{\mathrm{non-T}}$   & Offset & 45$^{\circ}$ & 1pc  & Non-Turb\\
B$_{\mathrm{T}}$       & Offset & 45$^{\circ}$ & 1pc  & Turb\\
\hline
C$_{\mathrm{T}}$       & Offset & 0$^{\circ}$  & 1pc  & Turb \\
\hline
D$_{\mathrm{non-T}}$   & Offset & 0$^{\circ}$ & 1.5pc  & Non-Turb \\
\end{tabular}
\end{table}


\section{Results} 
\label{res}

{As described in Section~\ref{obser}, the Serpens Main cluster is divided into two sub-clusters, the SE and NW sub-clusters. All the observational evidence tends to show that the SE sub-cluster is at the interface of a cloud-cloud collision, as opposed to the much more quiescent NW sub-cluster. For this reason, in the cloud-cloud collision simulations we present here, we first attempt to reproduce the physical properties of the SE sub-cluster. However, given the obvious physical connection between these two sub-clusters we also try, in a second set of simulations, to trigger the formation of a NW-like sub-cluster, as a by-product of the collision. }


For comparison of the velocity structure of the simulations we will use the
observational data and results from \citet[][]{2010A&A...519A..27D}: IRAM 30m
telescope observations of C$^{18}$O J=1$\rightarrow$0. We also refer to the
analysis of Serpens molecular data from the JCMT GBS HARP data of C$^{18}$O
J=3$\rightarrow$2 \citep[][]{2010MNRAS...Graves}. For H$_{2}$ column density
comparisons, we used the SCUBA 850~$\mu$m emission from JCMT
\citep{1999MNRAS.309..141D}.

{The comparison of the simulations with the observations is primarily focused on observational characteristics of the Serpens SE sub-cluster given that this is the region which appears to be directly influenced by the collision. However, even in the short-cylinder runs which do not form the NW sub-cluster, some of the velocity characteristics of the entire cloud are also taken into account. Thus, the following characteristics are the main focus points for comparison}:

\label{parameters}

{
\begin{itemize}
\item Column density structure: an elongated/filamentary shape aligned in a NW-SE direction, sub-clumped (Fig.~\ref{fig:850micron}).
\item The mass and size of the southern clump compared with those of the denser parts of the SE sub-cluster ($\sim$30~M$_\odot$ within $\sim$0.025~pc$^{2}$, measured using the 850~$\mu$m emission above 190~mJy\,beam$^{-1}$, which corresponds to column densities above 0.1~g\,cm$^{-2}$ at a temperature of 10K).
\item Overall velocity structure: a single red-shifted component in the north, with a fainter blue component spatially offset to east (top panel of Fig.~\ref{fig:pvobs} top panels and Fig.~\ref{fig:bluered}). Double (overlapping) components on the south, where the sources are being formed (Fig.~\ref{fig:pvobs} lower panel and Fig.~\ref{fig:bluered}) with a gradient from east to west of increasing velocities, up to 1.5~kms$^{-1}$ (Fig.~\ref{fig:3c}). Moving further south, this velocity structure should evolve back into one component along the less dense gas of the filament (Fig.~\ref{fig:3c}, and lower PV diagram of Fig.~17 of \citet[][]{2010MNRAS...Graves}).
\item Velocity structure of less dense material: thin eastern filaments perpendicular to the main filament of Serpens (seen in blue and green on the left-hand side of the main filament in Fig.~\ref{fig:3c}).
\end{itemize}}

\begin{figure*}[!t]
\centering
\includegraphics[angle=270,width=0.63\textwidth]{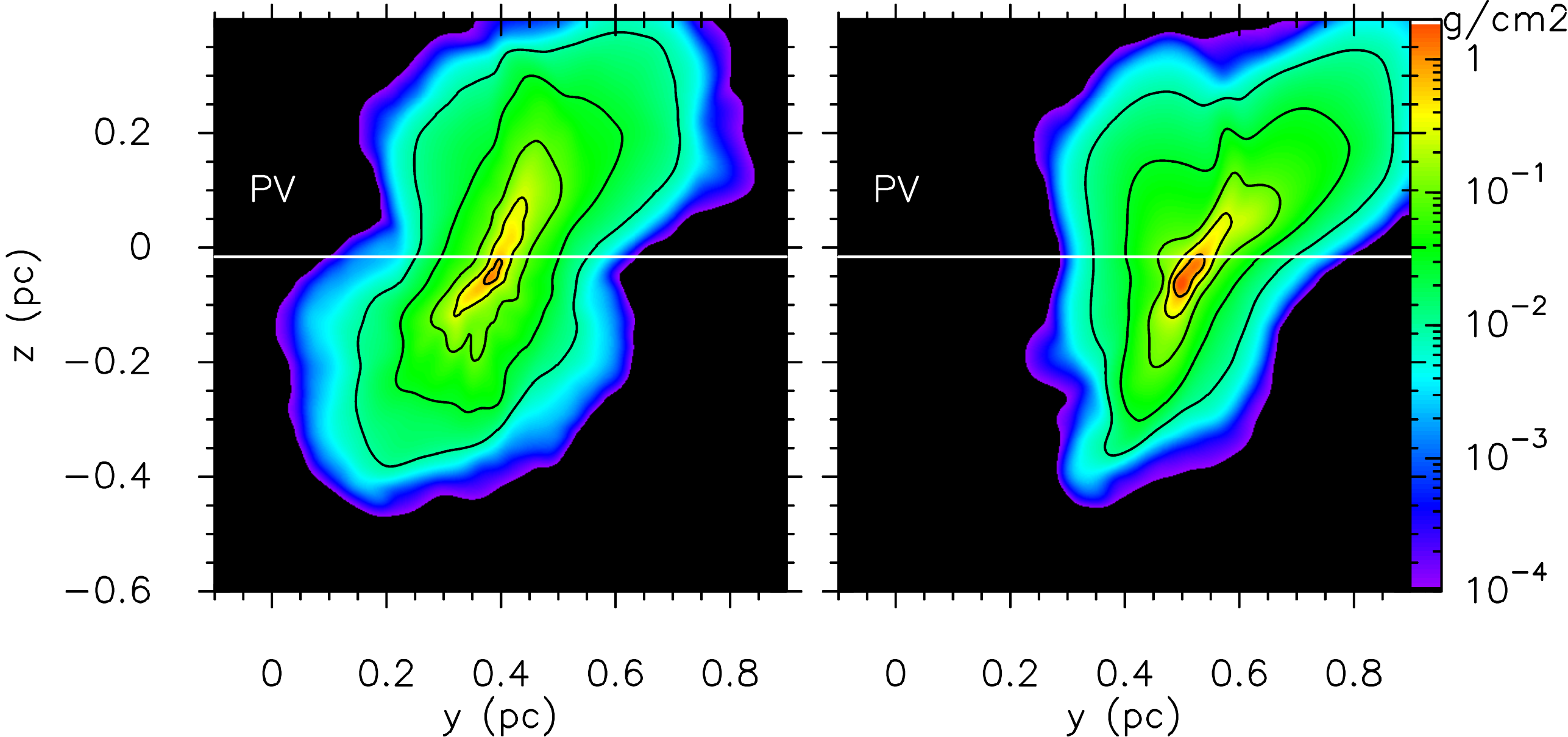}
\caption{\small{Colour scale and contours of the total column density along the line
  of sight, for the non-turbulent runs: the centered collision,
  A$_{\mathrm{non-T}}$, on the left; and off-centered collision,
  B$_{\mathrm{non-T}}$, on the right. The contours levels are 0.01, 0.03, 0.1,
  0.3 and 0.5 g~cm$^{-2}$ for both figures. {The white horizontal lines show the cuts for the PV diagrams in Fig.~\ref{fig:pvnonturb}}.}}
\label{fig:nonturb}
\end{figure*}

\begin{figure*}[!t]
\centering
\includegraphics[angle=270,width=0.63\textwidth]{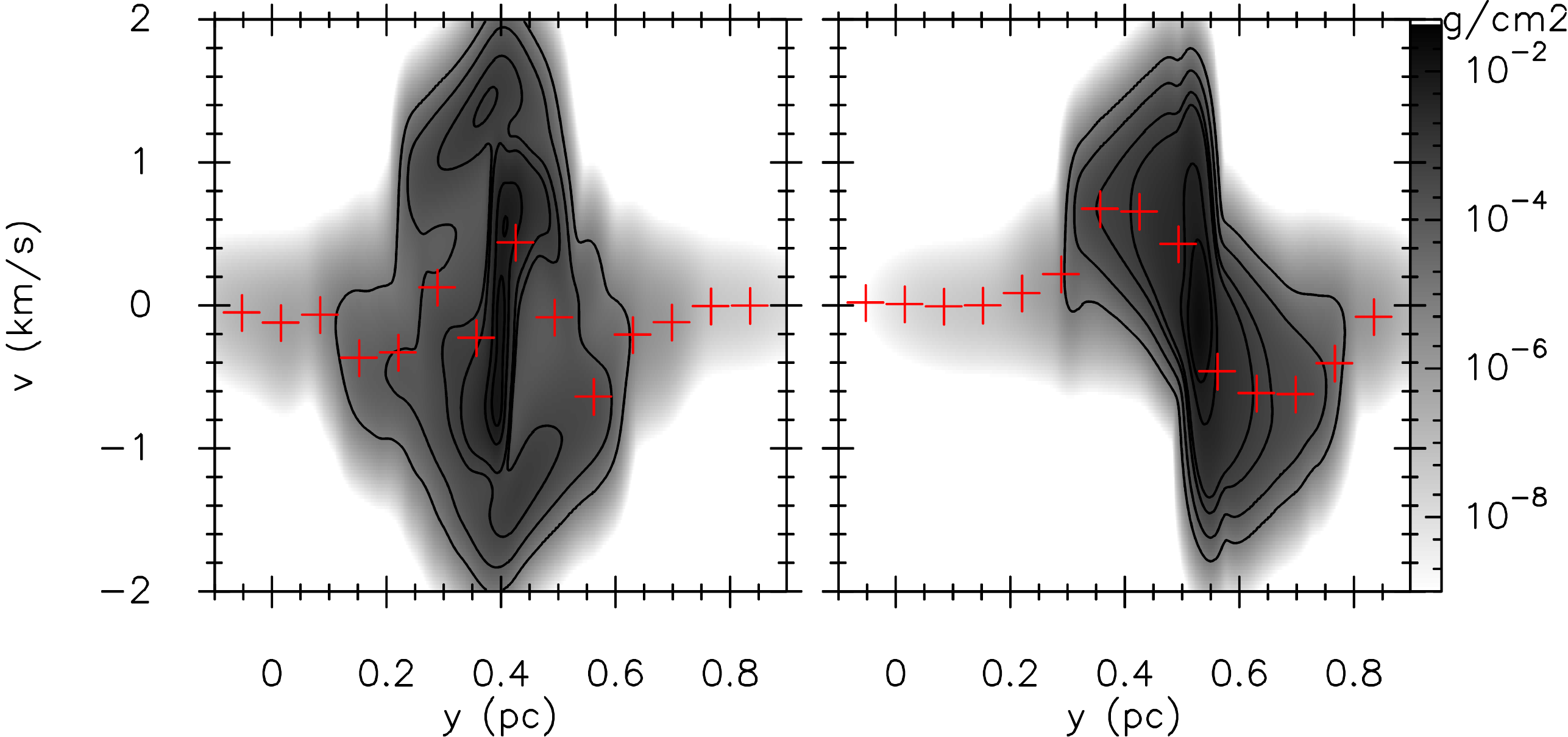}
\caption{\small{Position-velocity diagram at z~$=-0.02\:$pc (colour scale and contours) for the non-turbulent direct and offset runs, A$_{\mathrm{non-T}}$ and B$_{\mathrm{non-T}}$, left and right respectively. The {displayed velocity corresponds to the component along} the line of sight. Contours at 10$^{-5}$, 10$^{-4}$, 5~$\times~10^{-4}$, 10$^{-3}$, 5$\times10^{-3}$, 10$^{-2}$ and 3~$\times10^{-2}$~gcm$^{-2}$. The red crosses represent the column density weighted velocities, and are plotted as an auxiliary tool to see the velocity changes along the diagram. {Note that the velocities from the direct collision (left) in the central region seem to be more complex than those observed (Fig.~\ref{fig:pvobs} lower panel), whereas those from the offset collision (right) show a smooth shift in velocities from one component to another, likely because in the offset collision we see more clearly the parts of the clouds which do not collide.}}}
\label{fig:pvnonturb}
\end{figure*}	

In order to compare with the observations, we first constructed a
datacube from the simulated 3D cloud collisions. {For each simulation, we define the $x$
direction as the line of sight, so that the plane of the sky is the $yz$ plane. Choosing the time-step} of the
simulations where the first sink particle is formed, we created a
datacube of column density for a space-space-velocity 3D grid.  In the
spatial planes we convolved the models with a Gaussian of FWHM of 10 pix
(0.02~pc) which corresponds to the 22'' spatial resolution of the
IRAM-30m telescope observations of C$^{18}$O J=1$\rightarrow$0 at the 
distance of Serpens. While to reproduce the thermal velocity dispersion of Serpens, the velocity space was
convolved with a normalised Gaussian of 0.4~kms$^{-1}$ full width half maximum (FWHM),
correspondent to the thermal line width of H$_{2}$ at 10~K.

Gas temperature, density and abundance all affect the relationship
between the true column density of a cloud and the emission seen in a
molecular line. However as discussed in \citet[][]{2010A&A...519A..27D}
and \citet[][]{2010MNRAS...Graves} in Serpens the C$^{18}$O appears to be a
faithful tracer of the overall velocity structure of the cloud and not
significantly affected by outflow or infall motions, while globally
the 850~$\mu$m emission traces the mass distribution in the region.
To assess the success of the simulations in modeling Serpens, we
therefore compare the models with the velocity structure
of the C$^{18}$O and the overall column density distribution from the
dust emission.


\subsection{Modeling the Serpens SE sub-cluster}
\label{SE}

{We initially try to reproduce the characteristics of the SE sub-cluster, using the short-cylinders calculations: runs A, B
and C. In this section, we describe the results of these simulations. Taking the best-fit configuration for the short-cylinders, we then perform simulations with more extended cylinders designed to reproduce the whole cloud (Section~\ref{NWandSE}).}

\subsubsection{Non-Turbulent Runs: A$_{\mathrm{non-T}}$ and B$_{\mathrm{non-T}}$}
\label{nonT}

\begin{figure*}[!t]
\centering
\includegraphics[angle=270,width=0.63\textwidth]{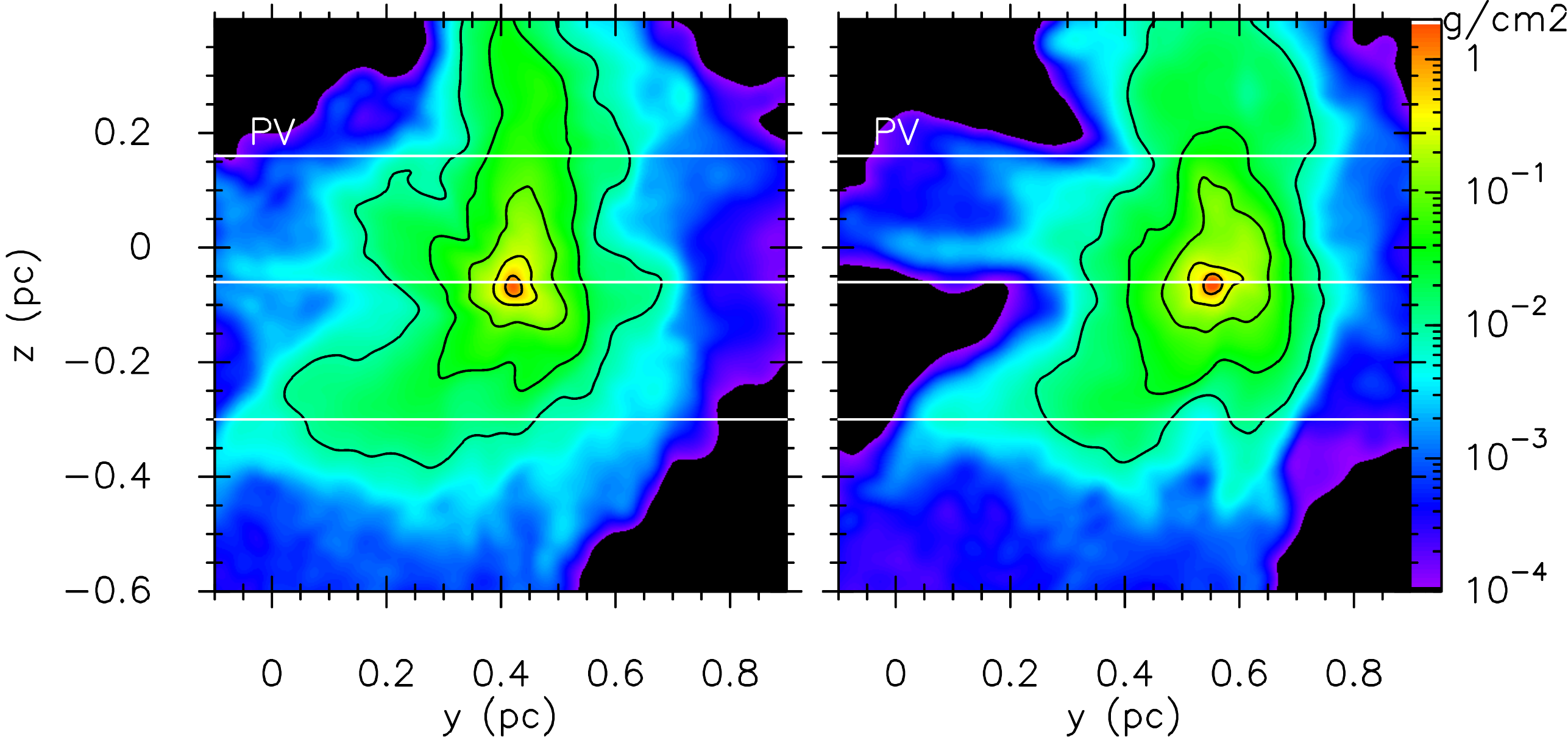}
\caption{\small{Colour scale and contours representing the total column density along
  the line of sight, for the turbulent runs: the centered collision,
  A$_{\mathrm{T}}$, on the left; and off-centered collision, B$_{\mathrm{T}}$,
  on the right. The contour levels are as in Fig.~\ref{fig:nonturb}. {The white horizontal lines show the cuts for the PV diagrams shown in Fig.~\ref{fig:pvturb}. Both turbulent runs show a final cloud which is much more structured better fitting the observations than Fig.~\ref{fig:nonturb}.}}}
\label{fig:turb}
\end{figure*}	

\begin{figure*}[!t]
\centering
\includegraphics[angle=270,width=0.63\textwidth]{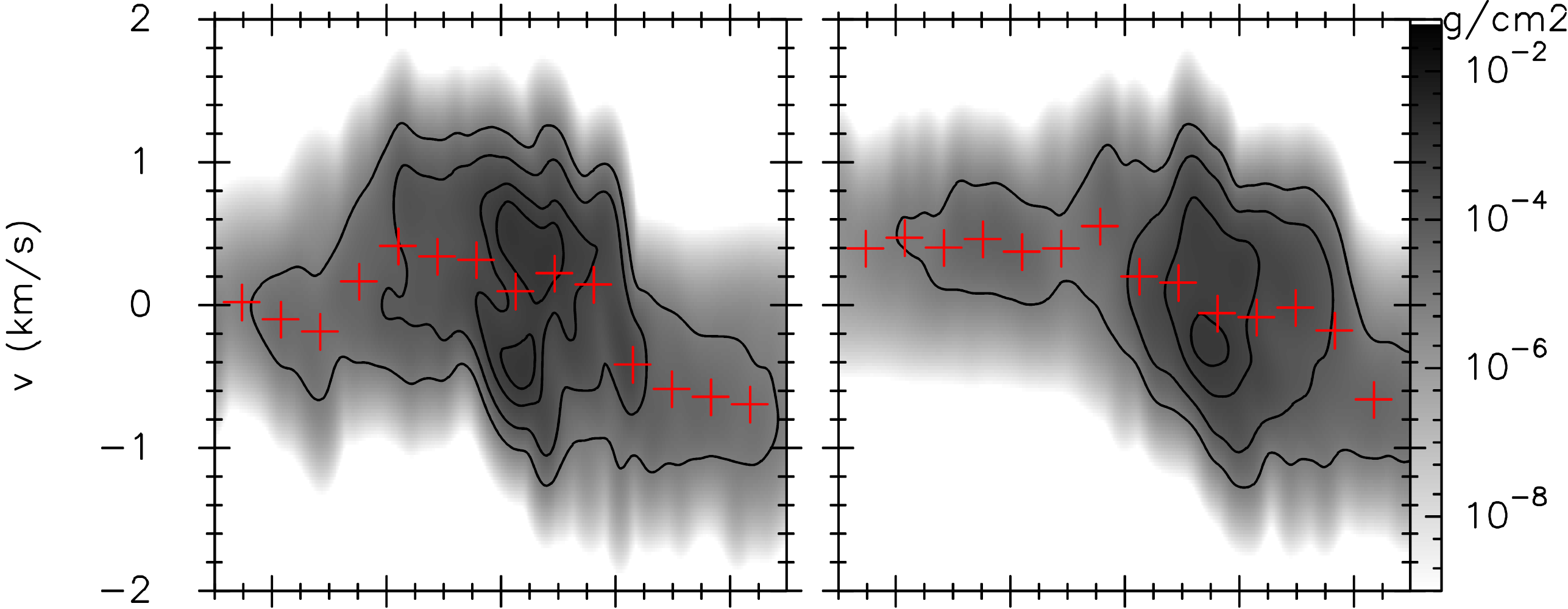}
\includegraphics[angle=270,width=0.63\textwidth]{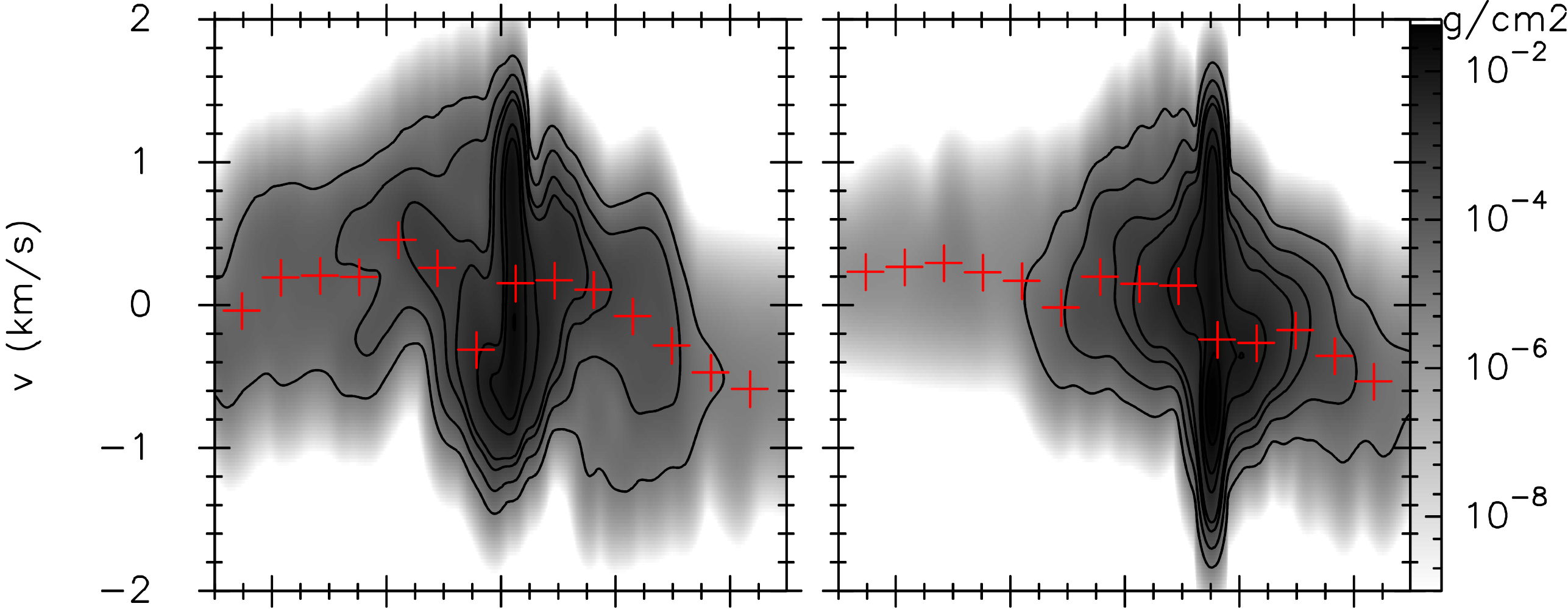}
\includegraphics[angle=270,width=0.63\textwidth]{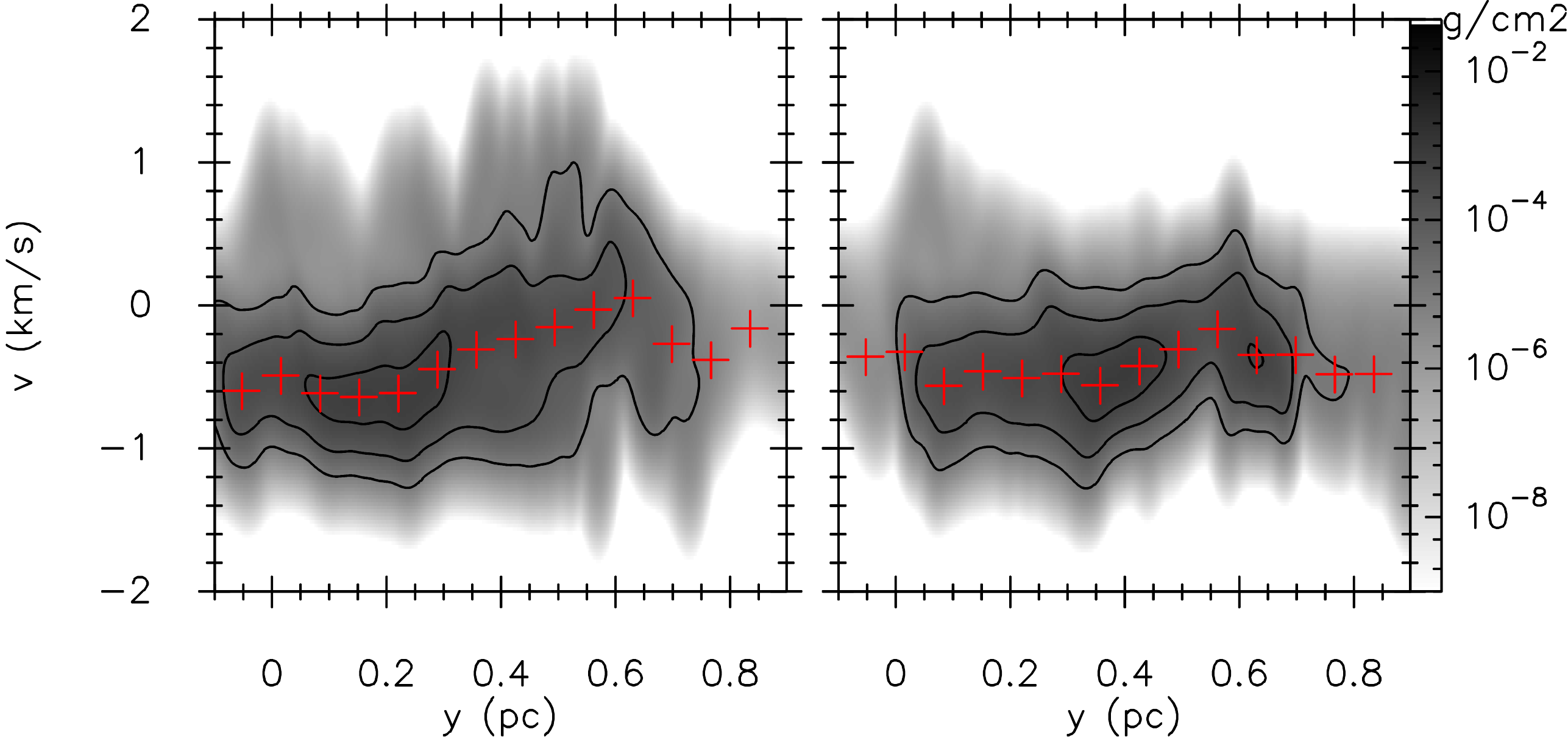}
\caption{\small{Position-velocity diagrams in colour scale and contours, at
  z~$=0.16$~pc (top), z~$=-0.06$~pc (middle) and z~$=-0.30$~pc (bottom) for the
  turbulent direct and offset runs, A$_{\mathrm{T}}$ and B$_{\mathrm{T}}$,
  left and right respectively. {The contour levels and red crosses are as in Fig.~\ref{fig:pvnonturb}. As expected from Fig.~\ref{fig:pvnonturb}, the velocities are more complex in the A run than those observed. The top panels show that the direct collision (left) produces two velocity components in the northern part of the cloud, whereas both the offset collision B (right) and the observations only show one (Fig.~\ref{fig:pvobs}). From the middle and lower panels, the differences are less obvious, but the middle panel shows that the direct collision shows a more ``messy'' velocity structure rather than a clear transition from high to low velocities.}}} 
 \label{fig:pvturb}
\end{figure*}
	
{Fundamentally, the only difference between these two simulations is that in the A$_{\mathrm{non-T}}$ case the centres of gravity of the cylinders are colliding head-on, maximising the loss of kinetic energy in the shock, while in the other case, B$_{\mathrm{non-T}}$, the centres of gravity are shifted and therefore the collision becomes softer, with the gas being able to conserve more of its initial kinetic energy. While the first configuration is a very peculiar one, the second one is very general, corresponding to the most likely situations of cloud-cloud collision.
As a result, the formation of sink particles in B$_{\mathrm{non-T}}$ is slower than in A$_{\mathrm{non-T}}$, and while it occurs only at the interface region for the latter, in B$_{\mathrm{non-T}}$ sink particles are more widely distributed, along a dense filament connecting both clouds (see Fig.~\ref{fig:nonturb} and App.~\ref{snapshots}, Fig.~\ref{fig:nonturb_time}).}

{The most significant difference between the results of these two simulations is the velocity structure at the collision interface. In A$_{\mathrm{non-T}}$, three distinct velocity components are present: the two velocities of the individual cylinder, plus an intermediate velocity of the shocked interface gas (Fig.~\ref{fig:pvnonturb} left panel). In B$_{\mathrm{non-T}}$, the velocity structure is continuously connecting the initial velocities of the cylinders (Fig.~\ref{fig:pvnonturb} right panel). Thus, the offset collision (B) is better able to reproduce the observations (Fig.~\ref{fig:pvobs}).}

\subsubsection{Turbulent Runs: A$_{\mathrm{T}}$ and B$_{\mathrm{T}}$}
\label{turb}

{In these simulations the net effect of turbulence is to render the cylinders inhomogeneous both in density and velocity, making them more realistic looking clouds. One of the important feature we see in Fig.~\ref{fig:turb} (and App.~\ref{snapshots}, , Fig.~\ref{fig:turb_time}), is the presence of channels of material, perpendicular to the main filament. These are reminiscent of features often observed in molecular clouds \citep[e.g.][]{2009ApJ...700.1609M}. These features will be further discussed in Sec.~\ref{discussion}. Another important aspect of injecting turbulence is that is generates density seeds. Before the first sink particles start to form, we see several regions where the density is starting to grow. As soon as one of these regions starts to be dense enough, it will attract the remaining less dense sub-clumps to form a major single structure.}



	
{At the end of both runs}, the material at a column density greater than 0.1~gcm$^{-2}$
is distributed over similar sized regions to the non-turbulent models,
0.025~pc$^{2}$ and 0.032~pc$^{2}$, for runs A$_{\mathrm{T}}$ and
B$_{\mathrm{T}}$ respectively. 
For A$_{\mathrm{T}}$, a total of 29~M$_{\odot}$ is in these 
high column density regions (39\% of the total mass of the cylinders), 
while for B$_{\mathrm{T}}$ the mass is 34~M$_{\odot}$ (45\% of the total mass).\\



{In terms of the velocity structure, 
the direct collision model, A$_{\mathrm{T}}$, still produces too many distinct velocity components with no smooth trends (Fig.~\ref{fig:pvturb} left panels), for the same reasons as for the A$_{\mathrm{non-T}}$ model. On the other hand, the B$_{\mathrm{T}}$ model produces a very interesting velocity structure. To the north (positive $z$)} we see mainly the tilted cylinder and we begin to detect the vertical cylinder on the left-hand side. The transition between the two is not sharp nor double peaked, but a rather broad smooth transition (Fig.~\ref{fig:pvturb} top right panel). A well defined double peak structure is only detected in the central region of the model where sink particles have formed (Fig.~\ref{fig:pvturb} middle right panel). South of this region, the column density becomes again dominated by one component from the tilted cylinder traveling away from us (Fig.~\ref{fig:pvturb} bottom right panel), as also found towards the south end of the Serpens filament by \citet[][]{2010MNRAS...Graves}. The results from this run (B$_{\mathrm{T}}$) are {more} consistent with the observations of the SE sub-cluster of Serpens (Sec.~\ref{data}), {but the change in velocity across the cloud is smaller than that observed ($\sim$0.5~kms$^{-1}$ instead of $\sim$1~kms$^{-1}$)}.

{We have also run a model with an intermediate turbulence level (not shown), with a 
velocity dispersion of 0.3~kms$^{-1}$. This model showed}
that the geometric configuration is the dominant factor in
determining the main velocity characteristics of the resulting cloud, while
the density distribution is sensitive to the level of turbulence. Lower levels
of turbulence result into a more filamentary structure, with more sink
particles forming along the denser filament. Increased values of initial
turbulence tend to disrupt the distribution of cloud's material, inhibiting
the formation of further clumps.


\subsubsection{Line of sight, short cylinder Run: C$_{\mathrm{T}}$}
\label{los_short}

Model C$_{\mathrm{T}}$ is simply a re-projection of model B$_{\mathrm{T}}$ so
that it is viewed along the axis of the relative motion of the cylinders.
{The time evolution of C$_{\mathrm{T}}$ is shown in App.~\ref{snapshots}, Fig.~\ref{fig:los_rot_time}.}

\begin{figure}[!t]
\centering
\includegraphics[angle=270,width=0.37\textwidth]{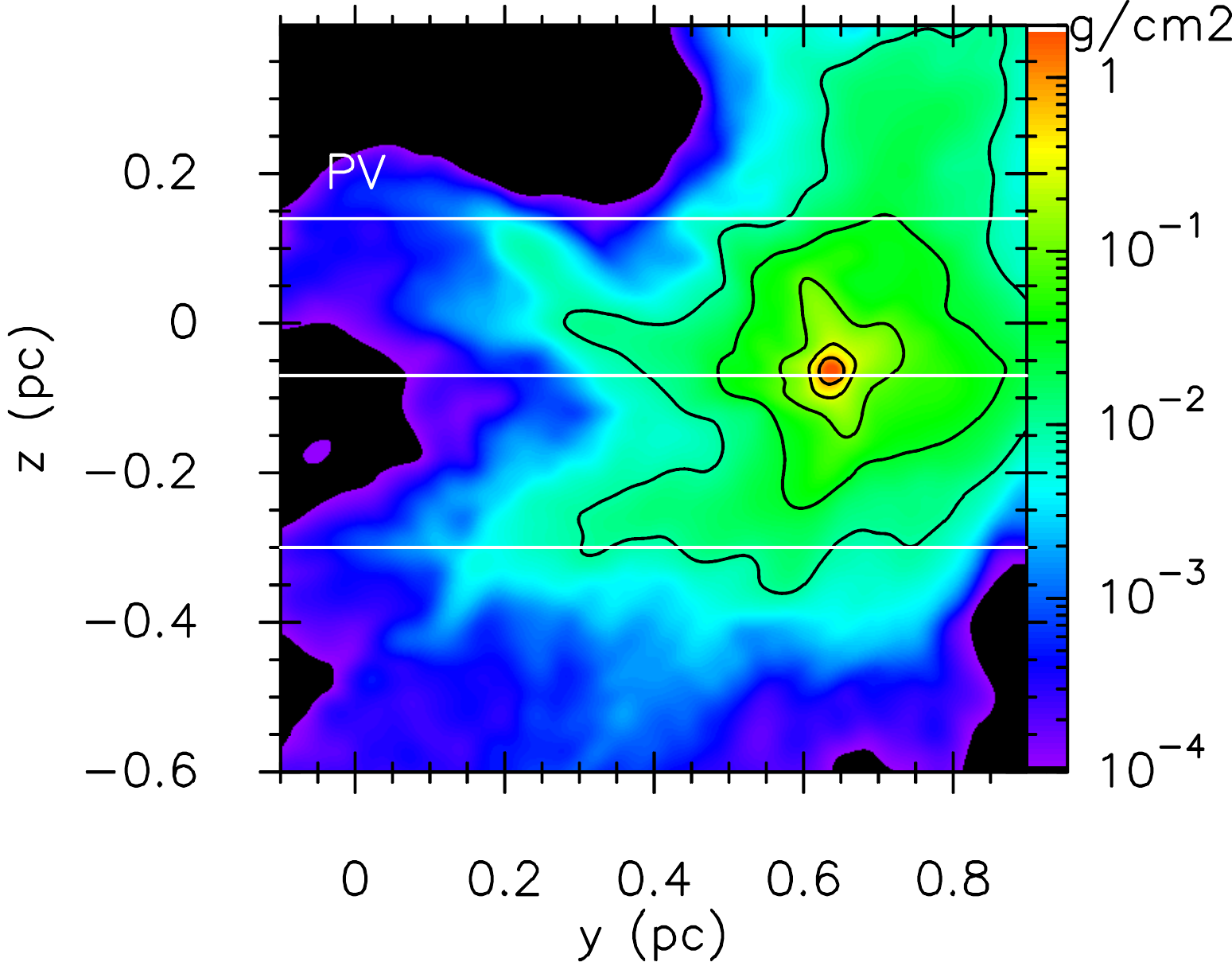}
\caption{\small{The total column density for the purely along the line of sight turbulent run, C$_{\mathrm{T}}$. Unlike
  runs A and B, we find a low density region in the north (large $\mathrm{z}$),
  as with this orientation we see regions which have not collided. 
  The contour levels are as in Fig.~\ref{fig:nonturb}.}}
\label{fig:cyllos}
\end{figure}

From this perspective (Fig.~\ref{fig:cyllos}) the low density material is less filamentary and more
spatially extended, as the parts of the cylinders which do not interact do not
overlap along the line of sight any longer. Therefore, the extent of the
lower density material is greater to the north, where the cylinders do not
merge, than in the south {(negative $z$)}. In this projection, the material at high
column density above 0.1~gcm$^{-2}$ covers a region of 0.02pc$^{2}$ in area
with a mass of 22M$_{\odot}$.

\begin{figure}[!t]
\centering
\includegraphics[angle=270,width=0.38\textwidth]{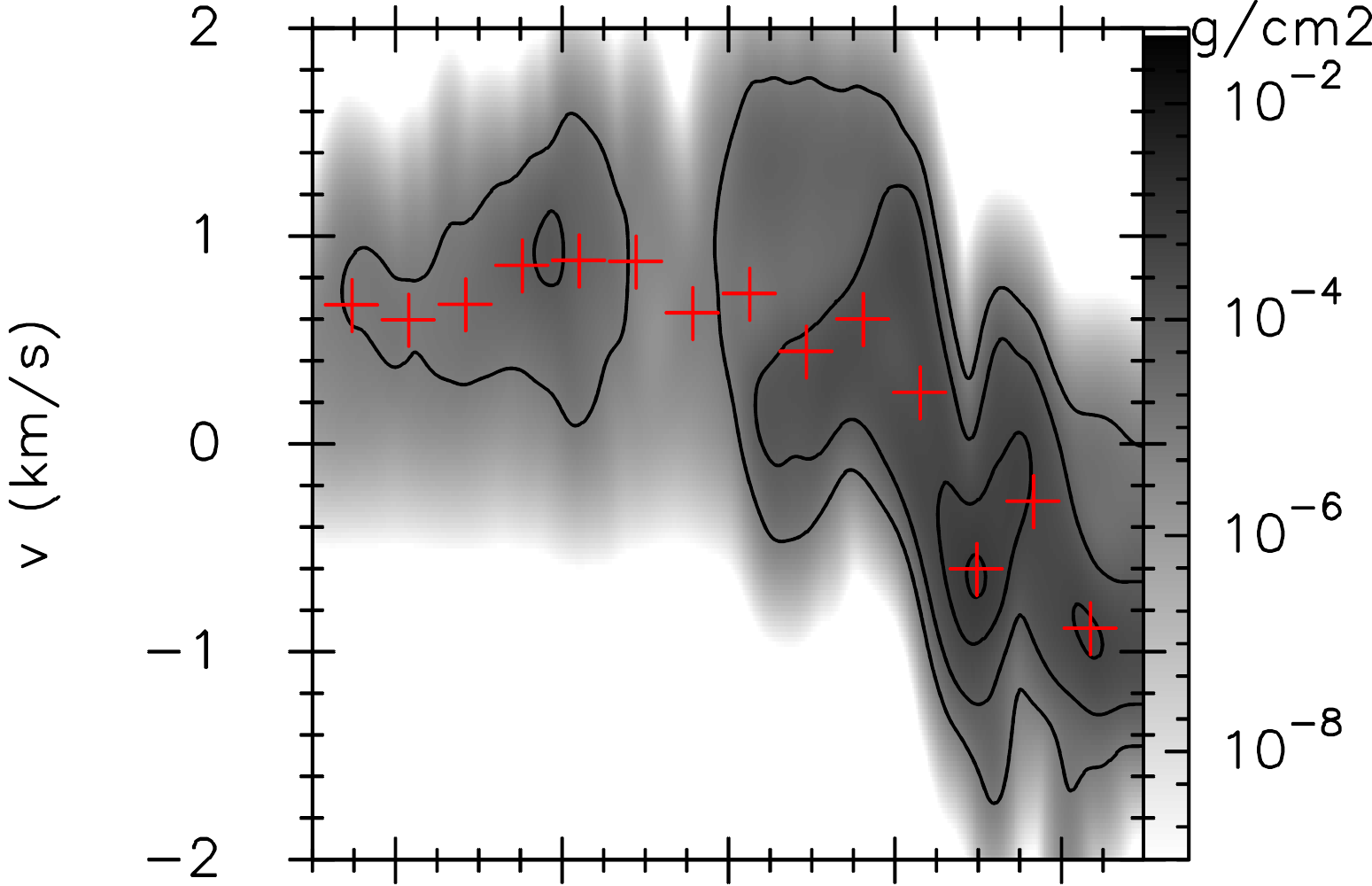}\\
\includegraphics[angle=270,width=0.38\textwidth]{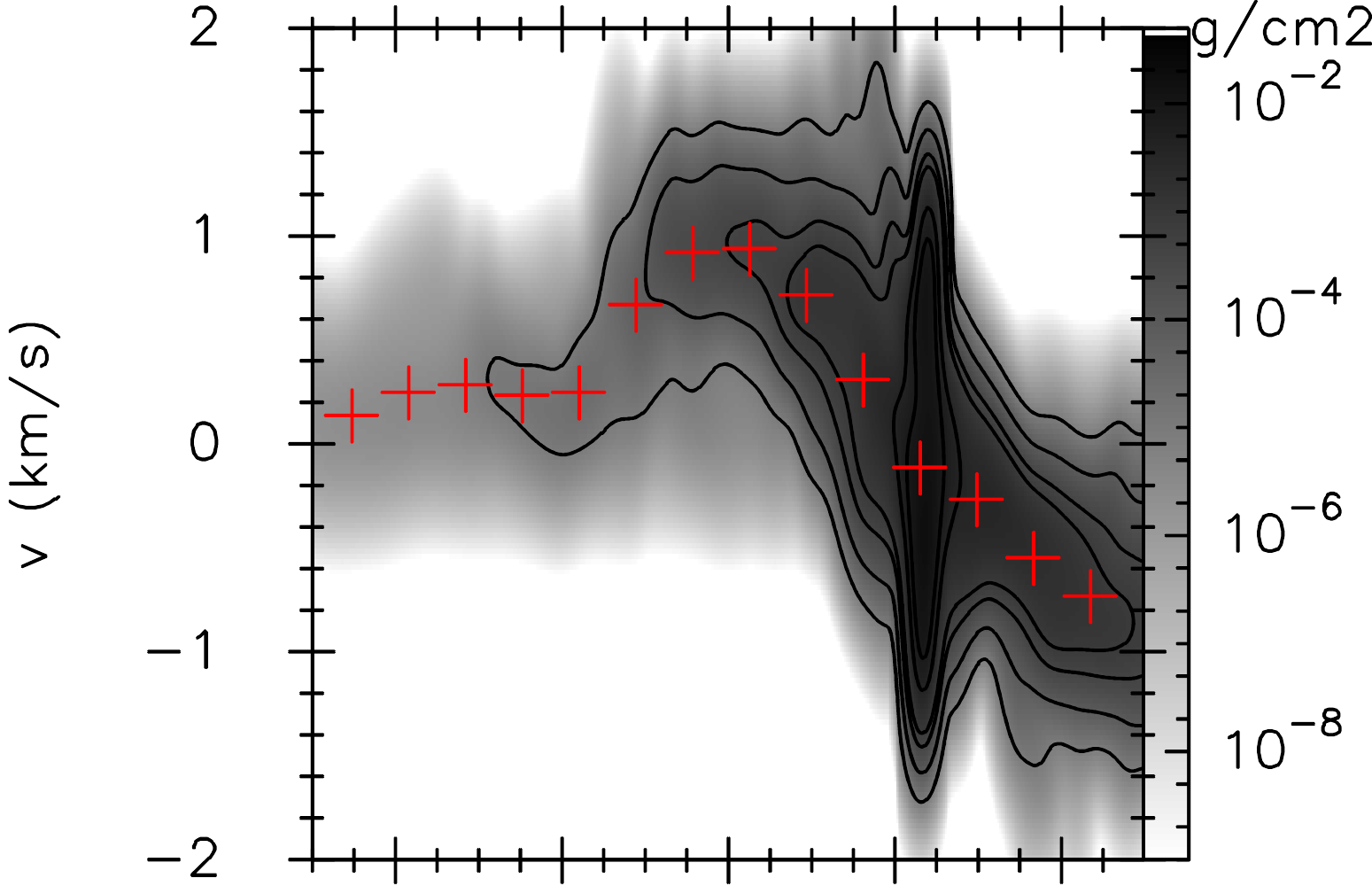}\\
\includegraphics[angle=270,width=0.38\textwidth]{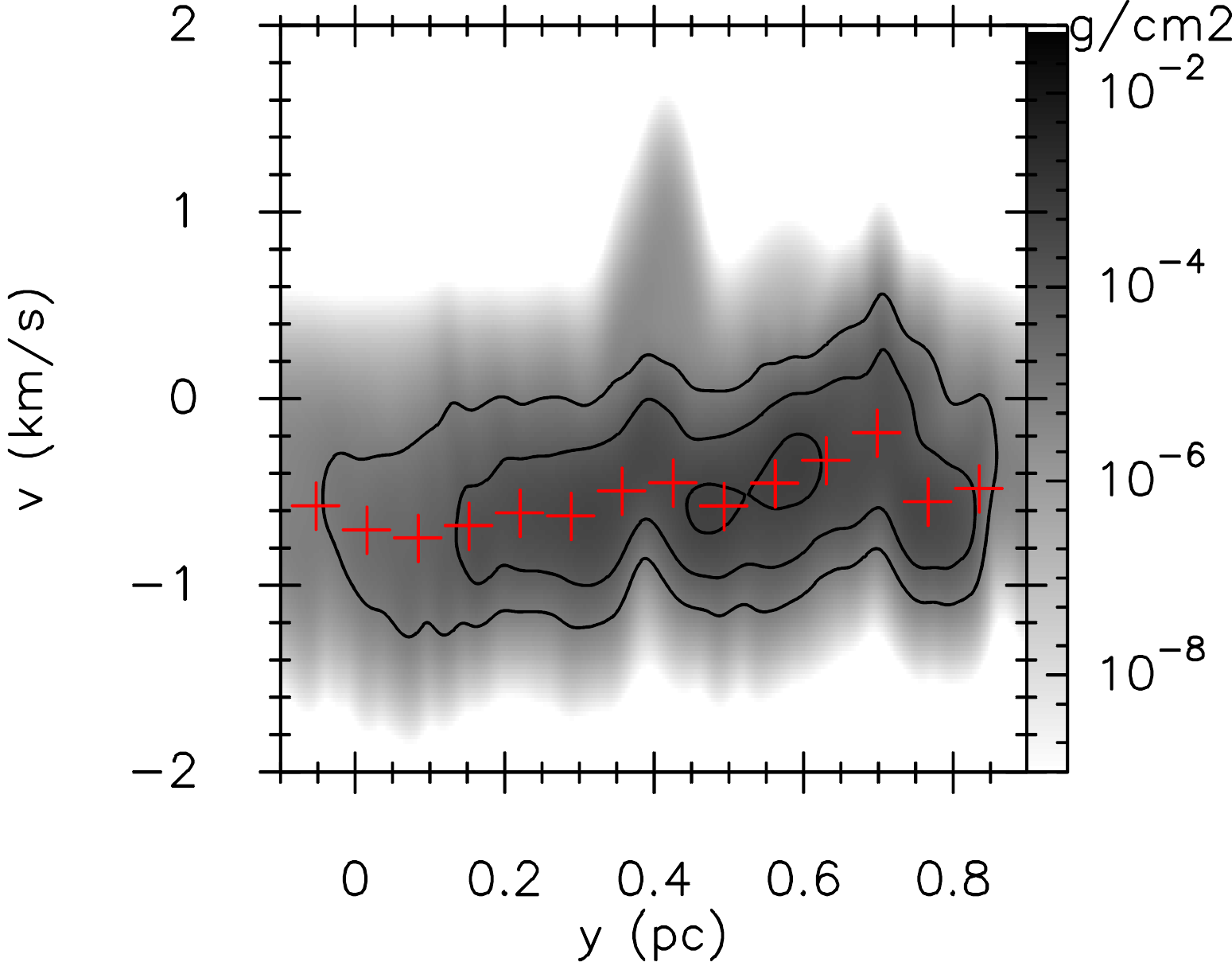}
\caption{\small{Position-velocity diagram in colour scale and contours, at
  z~$=0.14$~pc (top), z~$=-0.07$~pc (middle) and z~$=-0.30$~pc (bottom), for the
  turbulent, line of sight, short cylinders run C$_{\mathrm{T}}$.
{The contour levels and red crosses are as in Fig.~\ref{fig:pvnonturb}. Compared to model B$_{\mathrm{T}}$ (Fig.~\ref{fig:pvturb} right column), we see more evidence of the original velocities of the clouds from the non-interacting parts, due to the change of perspective. This calculation produces the best fit to the observations (Fig.~\ref{fig:pvobs}).}}}
\label{fig:pvlos}
\end{figure}

The general shape and trend of the velocity structure (Fig.~\ref{fig:pvlos}), although slightly more
complex, is similar to that seen in B$_{\mathrm{T}}$ and consistent with the velocity trend seen on the observational PV diagrams.
{Overall, this run (C) reproduces Serpens better than run B, as it now shows a more evident velocity change across the cloud.}


\subsection{Representation of both SE and NW Serpens sub-clusters}
\label{NWandSE}

	\subsubsection{Line of sight, long cylinders, Run: D$_{\mathrm{non-T}}$}
	\label{subsub:D}

Overall, model C$_{\mathrm{T}}$ satisfactorily reproduces the observations of
the SE sub-cluster. However it does not have enough Jeans masses to form a second
sub-cluster comparable to the NW sub-cluster in Serpens even though the
velocity structure of the simulation is very similar to that observed.
Therefore, we increased the length and mass of the cylinders to investigate
the possible formation of a separate structure in the north where the
cylinders do not interact. If the non-interacting northern region is massive
enough to be close to unstable, the collision in the south of the cylinders
may induce its rapid collapse without greatly affecting its systemic velocity.

\begin{figure}[!t]
\centering
\includegraphics[angle=270,width=0.37\textwidth]{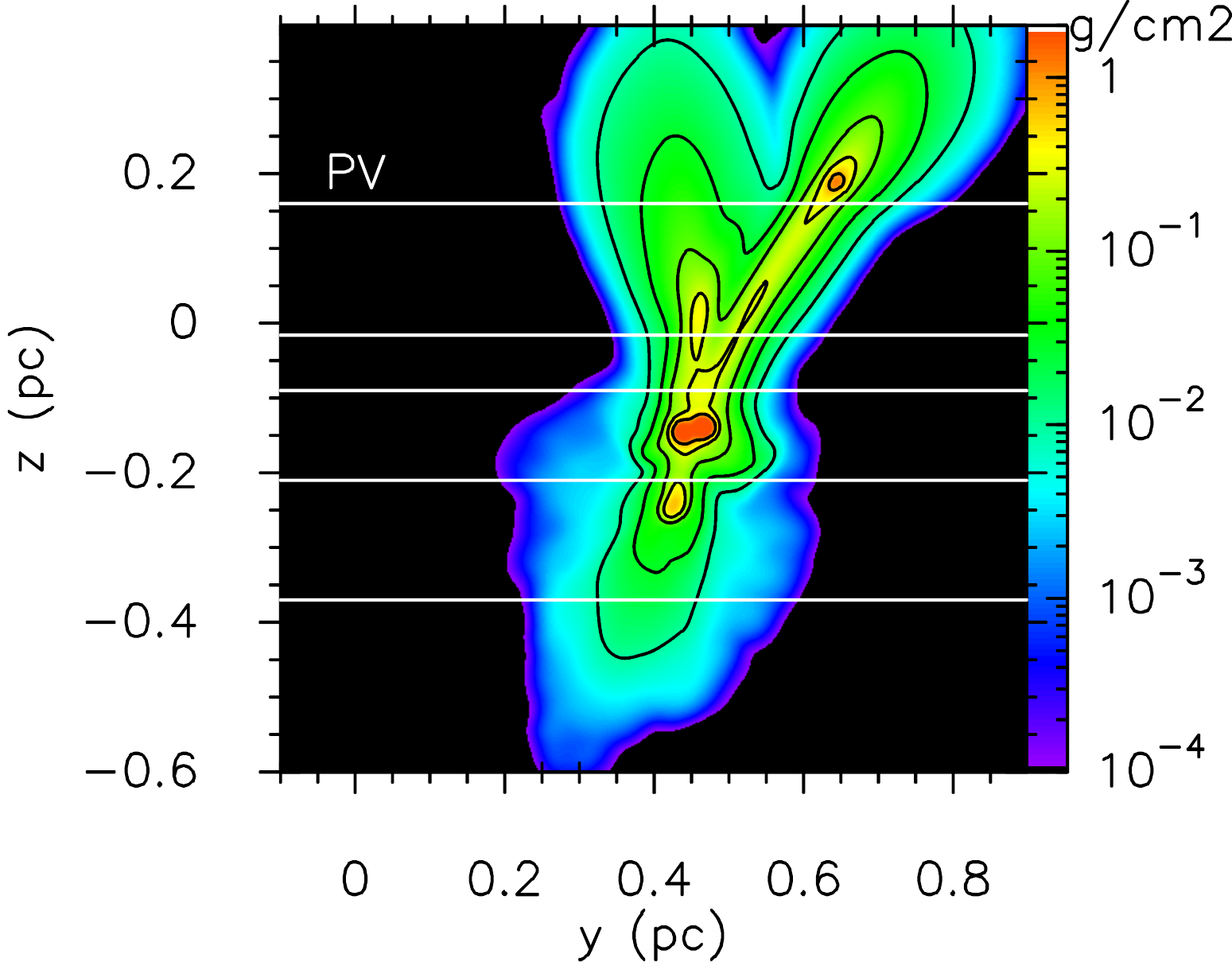}
\caption{\small{Total column density for the non-turbulent, purely along the line of
  sight run with longer cylinders, D$_{\mathrm{non-T}}$, {to test the production of both 
  the NW and SE sub-clusters of Serpens}. At this stage, sink
  particles have {indeed} formed where the collision is
  happening and in the north-west (within the tilted/longer cylinder). Given
  the geometry of the collision, the two cylinders do not end up colliding in
  the north. {However}, the only perspective where we do not see the cylinders
  overlap is the one we took for this run, where the line of sight is aligned
  with the motion of the cylinders. The contour levels are as in Fig.~\ref{fig:nonturb}.}}
\label{fig:cylloslong}
\end{figure}

A non-turbulent model with $\sim$1.5~pc long cylinders, colliding off-centre and along the
light of sight (model D$_{\mathrm{non-T}}$) does indeed form two
sub-clusters. The collision in the south leads to the formation of a
sink particle in the south, i.e. the synthetic SE sub-cluster. Later on, a sink
particle is formed in the north of the tilted cylinder, along the cylinder axis
({time evolution in App.~\ref{snapshots}, Fig.~\ref{fig:long_rot}}). Finally,
more sink particles are formed in the south, as the vertical cylinder crosses
through the tilted one (Fig.~\ref{fig:cylloslong}). At the end of this run, when sink particles are
formed throughout the clouds, the column density distribution is much more
filamentary than for the short cylinder calculation. There are two visible
sub-clusters: one in the NW and one in the SE. The total mass in these
sub-clusters amounts to 57~M$_{\odot}$, 52\% of the total mass in this case,
distributed over an area of 0.035~pc$^{2}$. However, the relative size and mass
of each individual sub-cluster is not quite as similar as seen in Serpens. The
simulated SE sub-cluster has a mass of about 30~M$_{\odot}$ in an area of
$\sim$0.015~pc$^{2}$, whereas the NW sub-cluster contains 12~M$_{\odot}$ in a
similar area.

The formation of a sink particle in the north occurs at $8\times10^{5}$yr and is mainly 
due to the initial instability of the cylinder itself. As a test, we ran a simulation with only
an isolated non-turbulent cylinder, with the same mass and size as the tilted
one. This simulation showed that it would form sink particles along the
cylinder axis at $10^{6}$yr. Therefore, the collision in the south does
not trigger the formation of this sink particle in the north, but only speeds
up its collapse.

Interestingly, several different turbulent runs with this configuration failed to induce the collapse of any structure in the north. The collapse only occurs with extremely low levels of turbulence (with amplitude of $\sim$ 0.05~kms$^{-1}$), almost indistinguishable from the non-turbulent run. This is because the turbulent velocities  cause the cloud to disperse from its initial configuration and { become less gravitationally bound than in the absence of turbulence}. From this, we conclude that the effect of the collision felt by the north region is quite subtle, and would only have an influence on the collapse there if the region is almost gravitationally unstable prior to the effect of any external perturbation.

The discrepancy in the relative masses of the two sub-clusters in the simulation compared to Serpens, and indeed the failure to form two sub-clusters in the turbulent simulations, could both be addressed by relaxing the over-simplistic uniform conditions in our initial conditions. Both the density distribution and turbulent velocity distributions are likely to be more inhomogeneous in real colliding clouds than in our models. {Also, this discrepancy could be suggesting that we are missing a physical ingredient such as magnetic pressure, which could help  maintaining the filamentary shape of the clouds even with high level of turbulence}. However we refrained from adding such additional complexity to the models as the nature of the inhomogeneities is poorly constrained and unlikely to provide any further significant physical insight into the processes in this region.

\begin{figure}[!t]
\centering
\includegraphics[angle=270,width=0.37\textwidth]{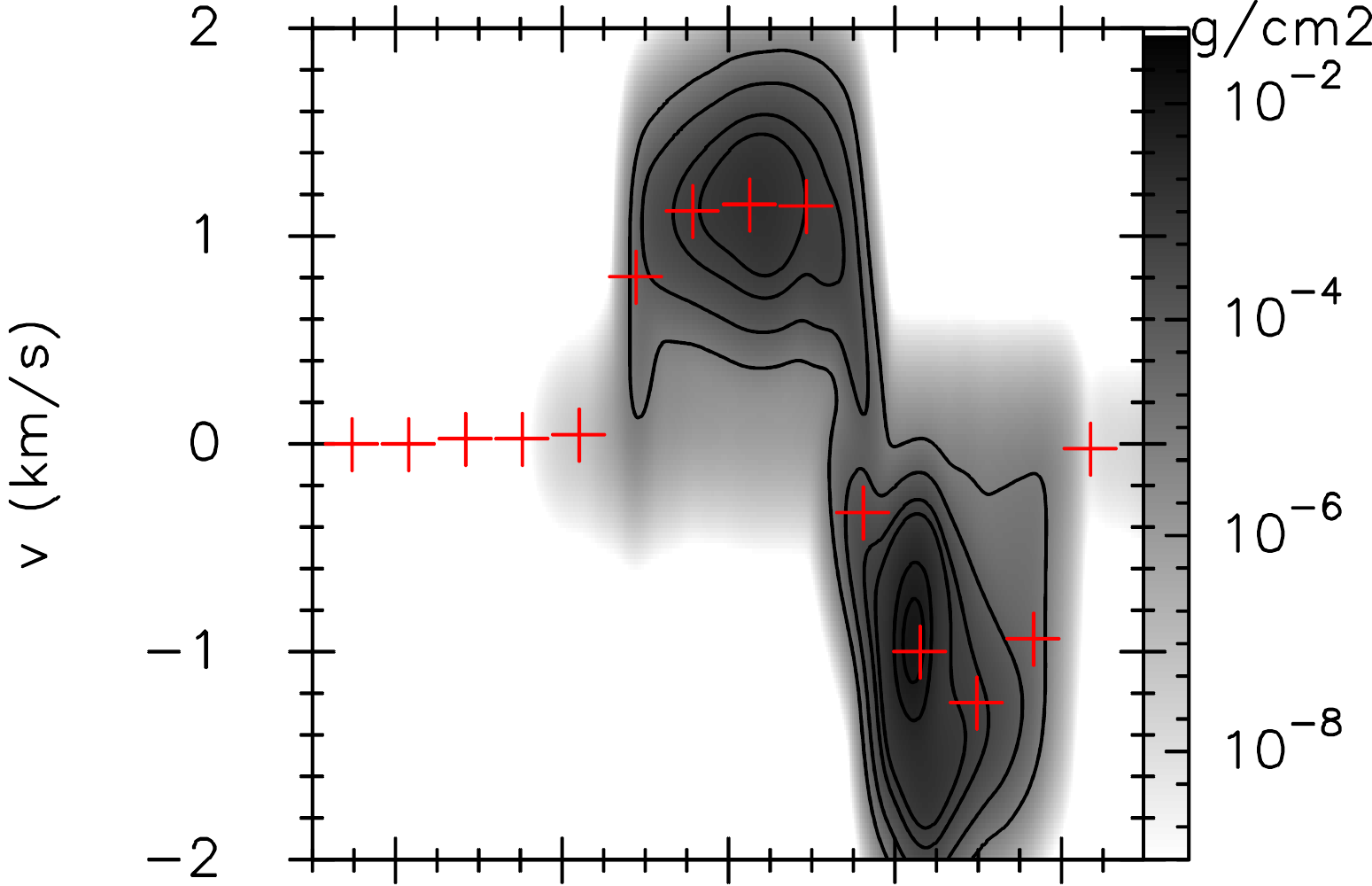}\\
\includegraphics[angle=270,width=0.37\textwidth]{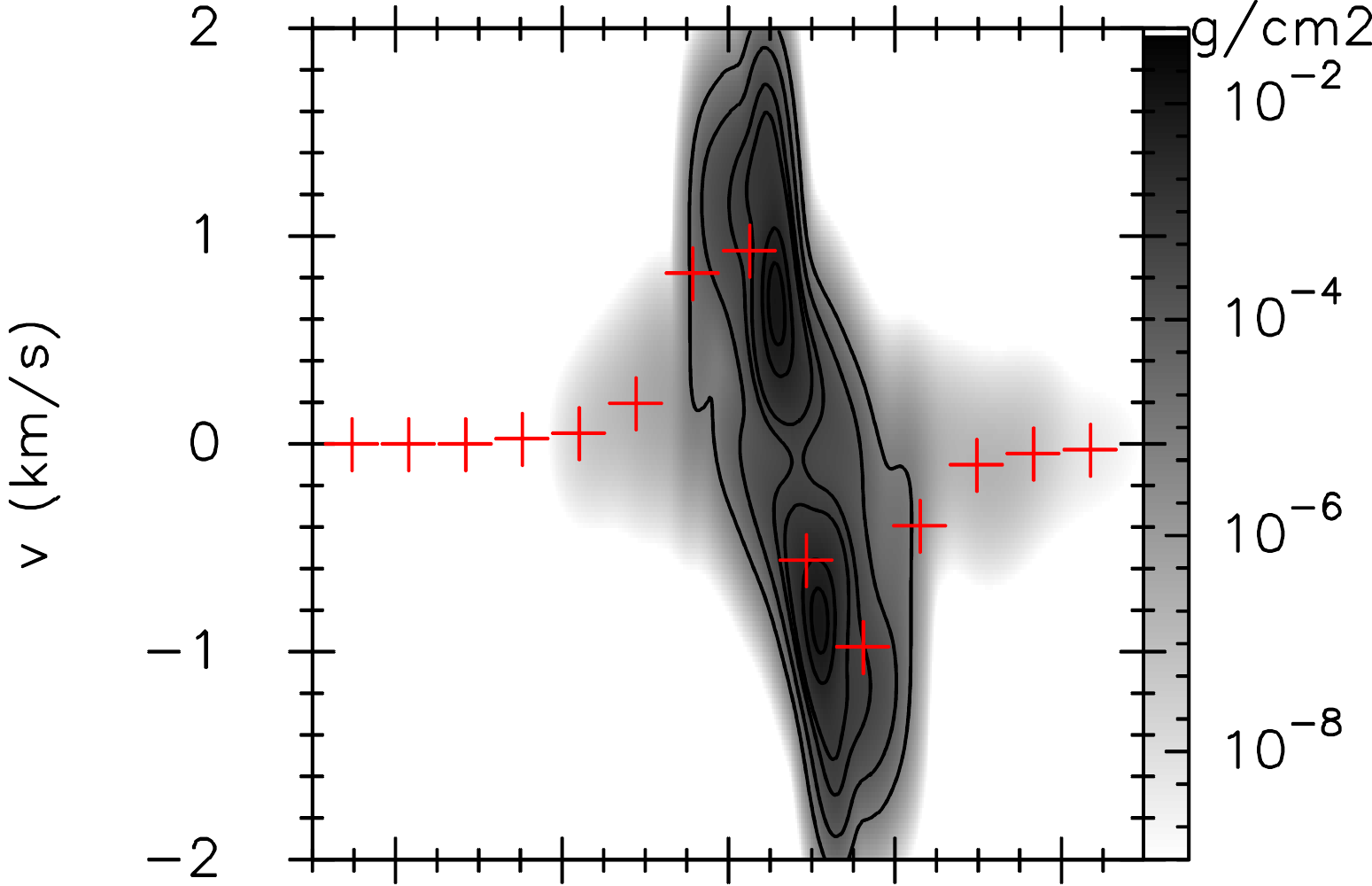}\\
\includegraphics[angle=270,width=0.37\textwidth]{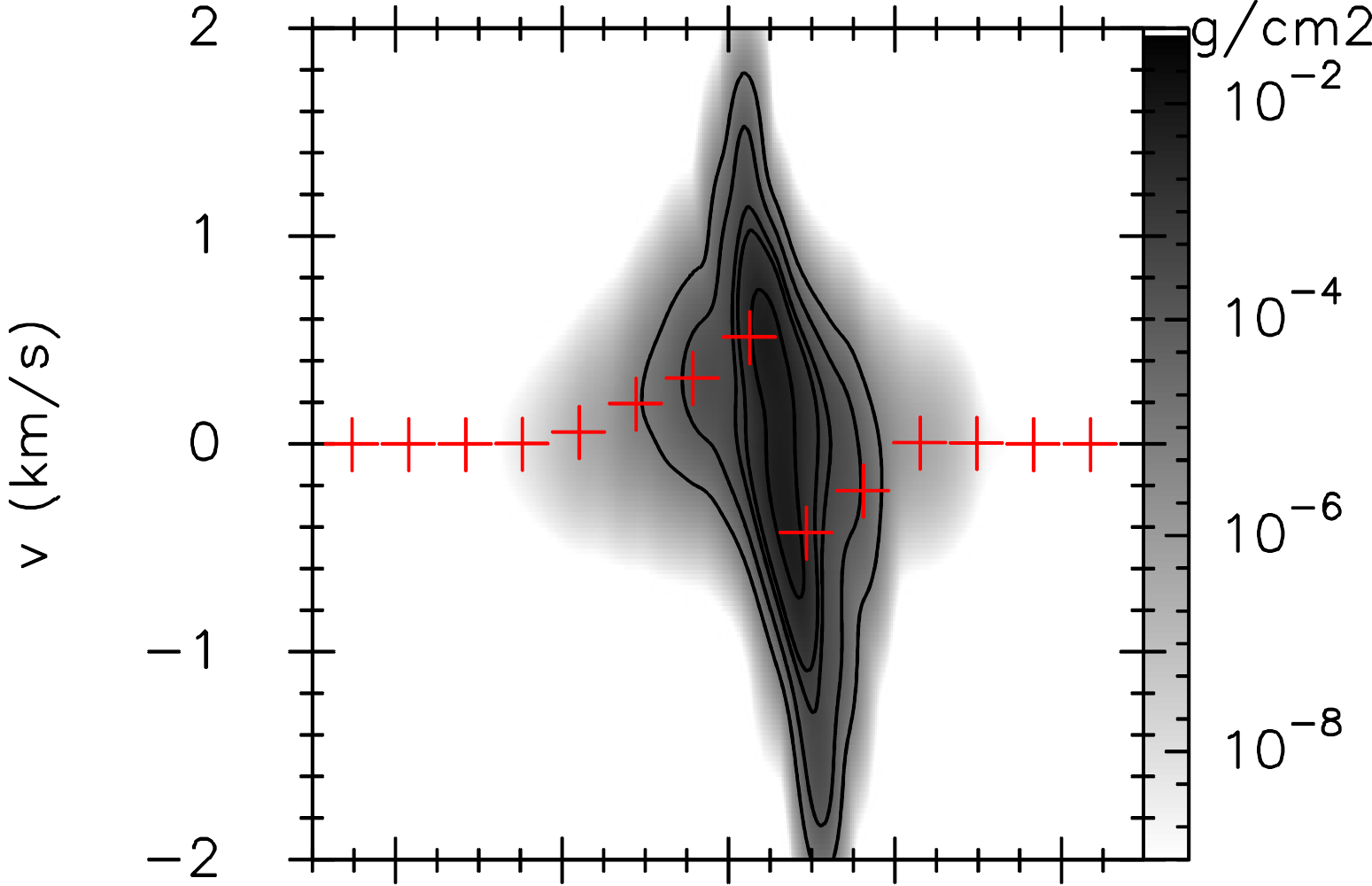}\\
\includegraphics[angle=270,width=0.37\textwidth]{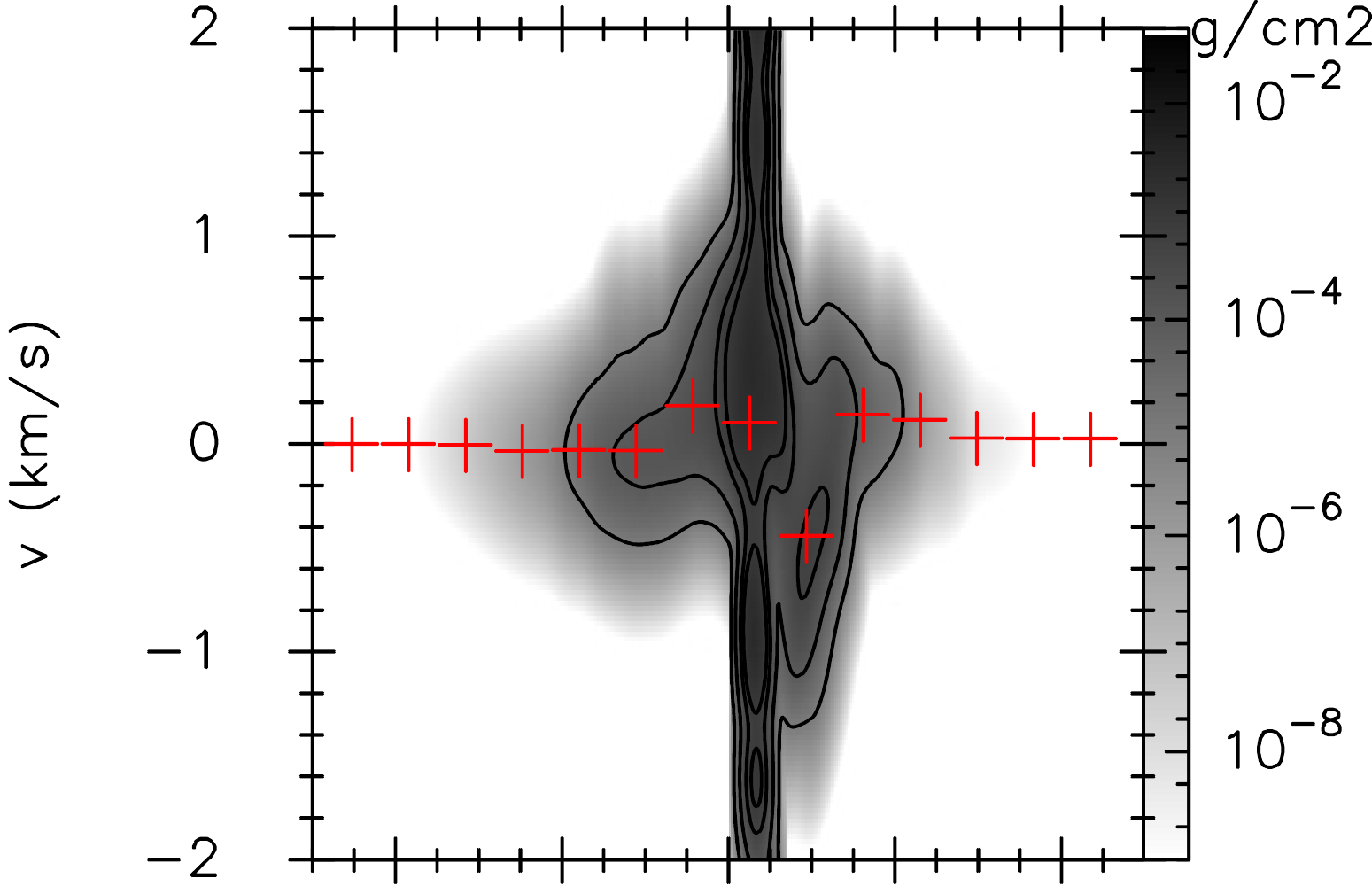}\\
\includegraphics[angle=270,width=0.37\textwidth]{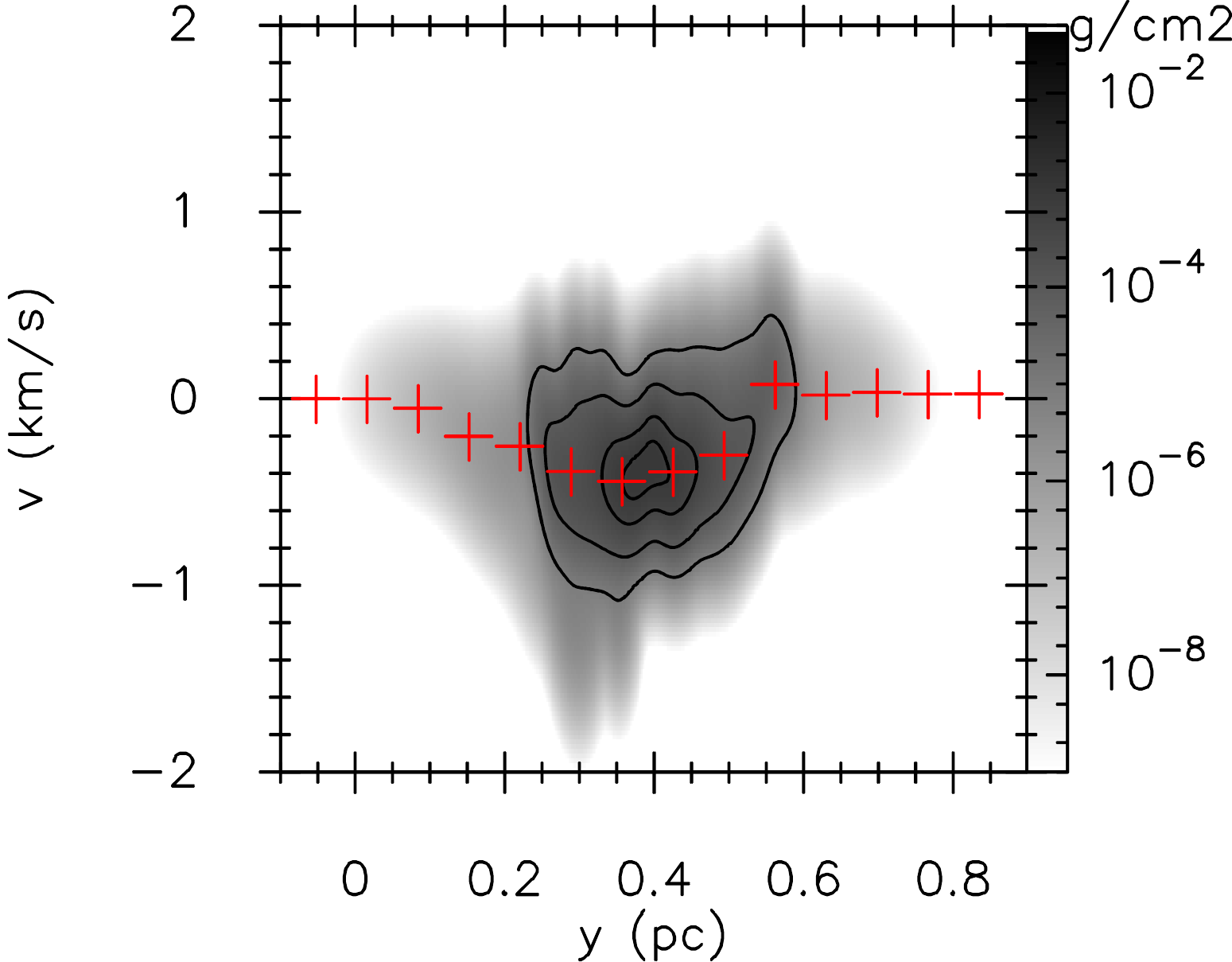}
\caption{\small{PV diagrams of the long cylinder run D$_{\mathrm{non-T}}$ at z~$=0.16$~pc, z~$=-0.02$~pc, z~$=-0.09$~pc, z~$=-0.21$~pc and z~$=-0.37$~pc (from top to bottom). {Contour levels and red crosses are as in Fig.~\ref{fig:pvnonturb}. {Despite the lack of structure typical from a non-turbulent case, these agree with the more detailed PV diagrams from observations from \citet{2010A&A...519A..27D} and \citet{2010MNRAS...Graves}.} }}}
\label{fig:pvloslong}
\end{figure}

In terms of its spatial evolution, the velocity structure of the D$_{\mathrm{non-T}}$ is similar to the observed one. {From Fig.~\ref{fig:pvloslong} we see that as we move from north to south (i.e. from positive $z$ to negative $z$), two spatially separated components, which then overlap where the collision takes place, to finally end-up with a single velocity component in the south. These PV diagrams can be compared to the more detailed PV diagrams across the entire cloud from \citet{2010A&A...519A..27D} and \citet{2010MNRAS...Graves}.}
Also note that the column densities are higher at velocities corresponding to the tilted cylinder. An increase in
the density of the tilted cylinder to produce a more massive NW sub-cluster would increase the difference in column densities between the two velocity components, more closely matching the observations. 



\subsection{Timescales}

{One interesting question for the star formation history of Serpens is to confirm whether or not the older population of pre-main sequence star of 2~Myr of age could be formed in the same cloud-cloud collision event. For this we need to estimate a number of characteristic timescales. First, all the simulations correspond to a total elapsed time of $\sim8 - 9 \times10^{5}$~years. However, it is only after about $2 - 3\times10^{5}$~years, about a third of the total simulation time, that the two cylinders start to interact. This time delay is needed, especially in the turbulent cases, to allow the cylinders to relax their initial density distributions. The time from the start of the interaction between the two clouds until they start forming sink particles is of the order of $6 \times 10^{5}$~years. For the 2~Myr stellar population to be able to form through this collision, the duration of the interaction time has to be at least 2~Myr. In fact, we would need the cloud to be an order of magnitude larger (i.e a few parsec wide) in order to still have an ongoing collision 2~Myr after the formation of the first protostars. This seems unrealistic, rather the cloud collision scenario we present for Serpens is consistent with the idea of two separate bursts of star formation in the region.}



\section{Discussion}
\label{discussion}

\begin{figure}[!t]
\centering
\includegraphics[width=0.4\textwidth]{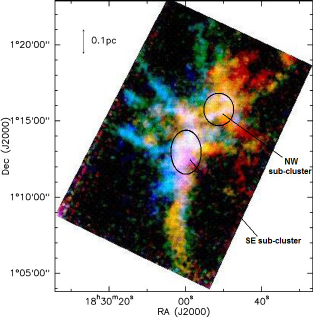}
\caption{\small{Velocity coded 3 colour plot of Serpens from the C$^{18}$O J=3$\rightarrow$2 data
  from the JCMT GBS. Each colour represents the maximum value in the velocity 
  intervals: blue: 5~$\rightarrow$~7.7~kms$^{-1}$; green: 7.7~$\rightarrow$~8.3~kms$^{-1}$; and red: 
  8.3~$\rightarrow$~11~kms$^{-1}$. Also discussed in \citet[][]{2010MNRAS...Graves}. The two sub-clusters of Serpens are indicated by the black ellipses.}}
\label{fig:3c}
\end{figure}



\begin{figure}[!t]
\centering
\vspace{0.5cm}
\includegraphics[width=0.48\textwidth]{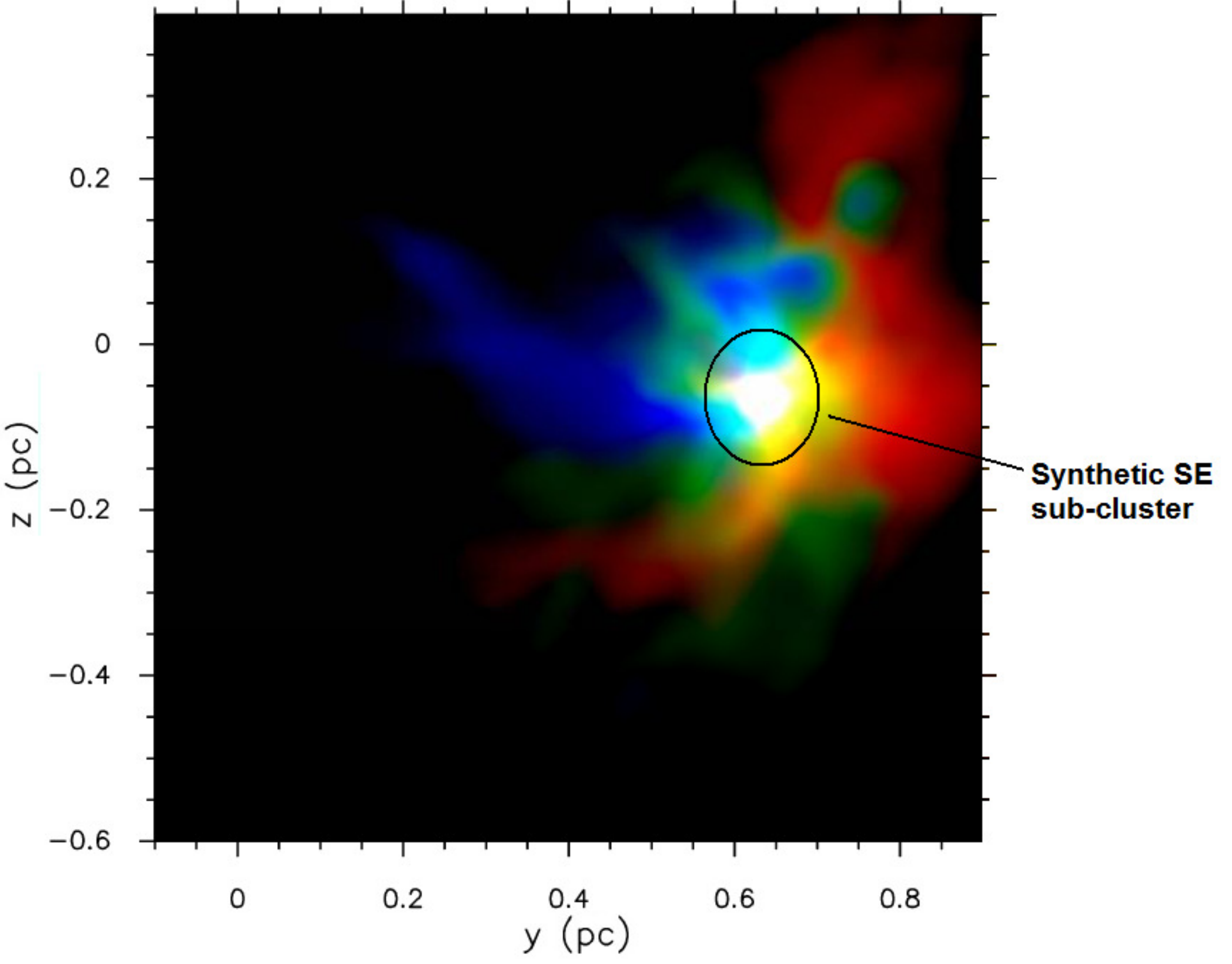}
\caption{\small{Velocity coded 3 colour plot of the simulation C$_{\mathrm{T}}$. Each colour represents the 
	total column density summed over the following line of sight velocities: blue: 0.3~$\rightarrow$~2~kms$^{-1}$; 
	green: -0.3~$\rightarrow$~0.3~kms$^{-1}$; and red: -2~$\rightarrow$~-0.3~kms$^{-1}$. This figure is to be compared with Fig.~\ref{fig:3c}, particularly around the SE sub-cluster. {The synthetic SE sub-cluster is indicated by a black ellipse.}}}
\label{fig:3colourC}
\end{figure}

Overall the simulation which best represents the SE sub-cluster of Serpens
star forming region is the model C$_{\mathrm{T}}$, the offset turbulent model
with short cylinders, reprojected so that the line of sight is coincident with
the direction of motion. It has a velocity structure similar to that observed, 
both comparing PV diagrams and the general trend on the velocity coded
3-colour plots (Fig.~\ref{fig:3c} and Fig.~\ref{fig:3colourC}). {Note that this simulation only represents the southern part of Serpens as the NW sub-cluster is not directly involved in the collision (cf Sec.~\ref{SE}). Despite this, the 3-colour plot of this run (Fig.~\ref{fig:3colourC}) shows a good agreement with that of the observations (Fig.~\ref{fig:3c}).} 
Model C$_{\mathrm{T}}$ shows an overall velocity gradient of $\sim$2~kms$^{-1}$ over 0.2 pc$^{-1}$, from blue in eastern regions to red on the west (Fig.~\ref{fig:3colourC}), similar to what is observed.

The 3-colour velocity figures also show the resemblance of the simulation C$_{\mathrm{T}}$ with the observations in the way that the filament extending south is mostly represented by the red/green velocities, while the blue part is mainly seen on the denser parts where the stars are being formed. The green-blue component is also seen further east, forming some less dense filaments perpendicular to the main filament. Note that we do not see these types of filament on the red side.

{Understanding the nature of these filaments can be important}. If these filaments had been caused by the initial
turbulence, then we would expect these to exist on both directions. {Alternatively, these filaments could be the consequence of some particles left behind as the cylinders move, resembling a tail.}
However, in simulation C, the cylinders' motion is along the line of sight, so if these
filaments resulted from a tail of material left behind during the collision
they would be behind {or in front of} the main structure. Their appearance therefore indicates
that these perpendicular filaments are caused by the geometry of the
collision, in particular, its asymmetric nature. These filaments place a
strong constraint to the geometry of the collision. The geometry of simulation
A (as well as other tests not shown) do not reproduce them.

In terms of its total column density distribution, 22~M$_{\odot}$ in
0.02~pc$^{2}$, model C closely matches the SE sub-cluster. Even the projected
distribution of material above a column density of 0.1~gcm$^{-2}$
resembles the shape of the SE sub-cluster.\\

To start to reproduce the NW sub-cluster in addition to the SE sub-cluster
requires the longer non-turbulent clouds used in model D$_{\mathrm{non-T}}$.
In this model a region close to being gravitationally
unstable can be perturbed and its collapse hastened by the collision, even
though it is not directly involved in the collision. This second sub-cluster
forms with the velocity of its native cloud and collapses smoothly and
independently of the southern region. 
If this model is allowed to run further, the NW sub-cluster falls onto the SE sub-cluster. 
Intriguingly, recent observations of the magnetic field in Serpens 
\citep[][]{2010ApJ...716..299S} appear to suggest the start of such a 
collapse which could result in the merging of the two sub-clusters. \\


\section{Conclusions}
\label{conc}

Serpens is a very interesting star forming region, not only for its youth, but
also for the striking differences between the two sub-clusters that compose
the active star-forming portion of the cloud. Even though they are at similar
stages of evolution, with most sources between Class 0 and Class I protostars
and similar dust continuum properties, the gas emission reveals that these
have not only different velocity characteristics but also different temperature
distributions \citep[][]{2010A&A...519A..27D}.

Motivated by the two velocity components seen in the southern sub-cluster of
Serpens, and the higher temperatures detected there, we performed several SPH
calculations of cloud-cloud collisions. The configurations used for these
simulations were based on the observed morphologies. 


For the SE sub-cluster, a model of two colliding clouds is able to reproduce both the column density structure, a centrally condensated filament aligned in a NW-SE direction, and the two velocity components seen where the star formation driven by the collision is occurring.  The same simulations did not, however, produce a second sub-cluster, similar to the NW sub-cluster of Serpens. Therefore, this sub-cluster does not seem to be the direct result of the collision. This was already suggested by the NW sub-cluster's well ``behaved'' temperature profile and velocity structure, as well as the uniform age of sources within the sub-cluster. However, the similar stage of evolution of the sources from the two sub-clusters and their proximity, suggests
that the two events are not totally independent.

A simulation with more elongated cylinders and increased masses provides a possible explanation. The presence of a marginally stable region in the northern part of one of the colliding clouds can have its collapse induced and quickened by perturbations driven by the cloud-cloud collision.

We consider a cloud-cloud collision scenario to be the best description of the driving of the star formation history in Serpens. Not only can it reproduce the observed velocities and column densities, as it offers a plausible explanation for why the two sub-clusters are so similar in some aspects and yet so different in others. Although cloud rotation may produce similar general velocity gradients to those observed, the complexity of the region is better explained with such a collision scenario, which is in essence similar to a shear-motion also suggested by \citet[][]{2002A&A...392.1053O}.

Despite the successful scenario provided by a cloud-cloud collision model, we failed to reproduce all of the Serpens characteristics in one single run. Additional support against gravity is required in order to sustain the existence of two different sub-clusters as in Serpens. The existing magnetic field of the region \citep[][]{2010ApJ...716..299S} could 
provide such a support.

\begin{verbatim}


\end{verbatim}

Ana Duarte Cabral is funded by the Funda{\c{c}}{\~a}o para a Ci{\^e}ncia e a Tecnologia of Portugal, under the grant reference SFRH/BD/36692/2007. A.D.C. would like to thank Jennifer Hatchell for useful discussions and suggestions that initially motivated this work. The calculations reported here were performed using the University of Exeter's SGI Altix ICE 8200 supercomputer. C.L.D.'s work was conducted as part of the award ``The formation of stars and planets: Radiation hydrodynamical and magnetohydrodynamical simulations'' made under the European Heads of Research Councils and European Science Foundation EURYI (European Young Investigator) Awards scheme and supported by funds from the Participating Organisations of EURYI and the EC Sixth Framework Programme.


\bibliographystyle{aa}	
\bibliography{references}		

\begin{thebibliography}{47}
\expandafter\ifx\csname natexlab\endcsname\relax\def\natexlab#1{#1}\fi

\bibitem[{{Anathpindika}(2009{\natexlab{a}})}]{2009A&A...504..451A}
{Anathpindika}, S. 2009{\natexlab{a}}, \aap, 504, 451

\bibitem[{{Anathpindika}(2009{\natexlab{b}})}]{2009A&A...504..437A}
{Anathpindika}, S. 2009{\natexlab{b}}, \aap, 504, 437

\bibitem[{{Andr{\'e}} {et~al.}(2010){Andr{\'e}}, {Men'shchikov}, {Bontemps},
  {K{\"o}nyves}, {Motte}, {Schneider}, {Didelon}, {Minier}, {Saraceno},
  {Ward-Thompson}, {di Francesco}, {White}, {Molinari}, {Testi}, {Abergel},
  {Griffin}, {Henning}, {Royer}, {Mer{\'{\i}}n}, {Vavrek}, {Attard},
  {Arzoumanian}, {Wilson}, {Ade}, {Aussel}, {Baluteau}, {Benedettini},
  {Bernard}, {Blommaert}, {Cambr{\'e}sy}, {Cox}, {di Giorgio}, {Hargrave},
  {Hennemann}, {Huang}, {Kirk}, {Krause}, {Launhardt}, {Leeks}, {Le Pennec},
  {Li}, {Martin}, {Maury}, {Olofsson}, {Omont}, {Peretto}, {Pezzuto}, {Prusti},
  {Roussel}, {Russeil}, {Sauvage}, {Sibthorpe}, {Sicilia-Aguilar}, {Spinoglio},
  {Waelkens}, {Woodcraft}, \& {Zavagno}}]{2010A&A...518L.102A}
{Andr{\'e}}, P., {Men'shchikov}, A., {Bontemps}, S., {et~al.} 2010, \aap, 518,
  L102+

\bibitem[{{Ballesteros-Paredes} {et~al.}(1999){Ballesteros-Paredes},
  {Hartmann}, \& {V{\'a}zquez-Semadeni}}]{1999ApJ...527..285B}
{Ballesteros-Paredes}, J., {Hartmann}, L., \& {V{\'a}zquez-Semadeni}, E. 1999,
  \apj, 527, 285

\bibitem[{{Bastien}(1983)}]{1983A&A...119..109B}
{Bastien}, P. 1983, \aap, 119, 109

\bibitem[{{Bate}(1995)}]{1995PhDT.......181B}
{Bate}, M. 1995, PhD thesis, PhD thesis, Univ.~Cambridge, (1995)

\bibitem[{{Bate}(2009{\natexlab{a}})}]{2009MNRAS.392..590B}
{Bate}, M.~R. 2009{\natexlab{a}}, \mnras, 392, 590

\bibitem[{{Bate}(2009{\natexlab{b}})}]{2009MNRAS.397..232B}
{Bate}, M.~R. 2009{\natexlab{b}}, \mnras, 397, 232

\bibitem[{{Bate} {et~al.}(2002){Bate}, {Bonnell}, \&
  {Bromm}}]{2002MNRAS.332L..65B}
{Bate}, M.~R., {Bonnell}, I.~A., \& {Bromm}, V. 2002, \mnras, 332, L65

\bibitem[{{Bate} {et~al.}(2003){Bate}, {Bonnell}, \&
  {Bromm}}]{2003MNRAS.339..577B}
{Bate}, M.~R., {Bonnell}, I.~A., \& {Bromm}, V. 2003, \mnras, 339, 577

\bibitem[{{Benz} {et~al.}(1990){Benz}, {Cameron}, {Press}, \&
  {Bowers}}]{1990ApJ...348..647B}
{Benz}, W., {Cameron}, A.~G.~W., {Press}, W.~H., \& {Bowers}, R.~L. 1990, \apj,
  348, 647

\bibitem[{{Dale} {et~al.}(2007){Dale}, {Bonnell}, \&
  {Whitworth}}]{2007MNRAS.375.1291D}
{Dale}, J.~E., {Bonnell}, I.~A., \& {Whitworth}, A.~P. 2007, \mnras, 375, 1291

\bibitem[{{Dale} {et~al.}(2009){Dale}, {W{\"u}nsch}, {Whitworth}, \& {Palou{\v
  s}}}]{2009MNRAS.398.1537D}
{Dale}, J.~E., {W{\"u}nsch}, R., {Whitworth}, A., \& {Palou{\v s}}, J. 2009,
  \mnras, 398, 1537

\bibitem[{{Davis} {et~al.}(1999){Davis}, {Matthews}, {Ray}, {Dent}, \&
  {Richer}}]{1999MNRAS.309..141D}
{Davis}, C.~J., {Matthews}, H.~E., {Ray}, T.~P., {Dent}, W.~R.~F., \& {Richer},
  J.~S. 1999, \mnras, 309, 141

\bibitem[{{Dobbs}(2008)}]{2008MNRAS.391..844D}
{Dobbs}, C.~L. 2008, \mnras, 391, 844

\bibitem[{{Dobbs} {et~al.}(2006){Dobbs}, {Bonnell}, \&
  {Pringle}}]{2006MNRAS.371.1663D}
{Dobbs}, C.~L., {Bonnell}, I.~A., \& {Pringle}, J.~E. 2006, \mnras, 371, 1663

\bibitem[{{Duarte-Cabral} {et~al.}(2010){Duarte-Cabral}, {Fuller}, {Peretto},
  {Hatchell}, {Ladd}, {Buckle}, {Richer}, \& {Graves}}]{2010A&A...519A..27D}
{Duarte-Cabral}, A., {Fuller}, G.~A., {Peretto}, N., {et~al.} 2010, \aap, 519,
  A27+

\bibitem[{{Dubinski} {et~al.}(1995){Dubinski}, {Narayan}, \&
  {Phillips}}]{1995ApJ...448..226D}
{Dubinski}, J., {Narayan}, R., \& {Phillips}, T.~G. 1995, \apj, 448, 226

\bibitem[{{Elmegreen} \& {Lada}(1977)}]{1977ApJ...214..725E}
{Elmegreen}, B.~G. \& {Lada}, C.~J. 1977, \apj, 214, 725

\bibitem[{{Galv{\'a}n-Madrid} {et~al.}(2010){Galv{\'a}n-Madrid}, {Zhang},
  {Keto}, {Ho}, {Zapata}, {Rodr{\'{\i}}guez}, {Pineda}, \&
  {V{\'a}zquez-Semadeni}}]{2010arXiv1004.2466G}
{Galv{\'a}n-Madrid}, R., {Zhang}, Q., {Keto}, E., {et~al.} 2010, ArXiv e-prints

\bibitem[{{Gittins} {et~al.}(2003){Gittins}, {Clarke}, \&
  {Bate}}]{2003MNRAS.340..841G}
{Gittins}, D.~M., {Clarke}, C.~J., \& {Bate}, M.~R. 2003, \mnras, 340, 841

\bibitem[{{Graves} {et~al.}(2010){Graves}, {Richer}, {Buckle}, {Duarte-Cabral},
  {Fuller}, {Hogerheijde}, {Owen}, {Brunt}, {Butner}, {Cavanagh},
  {Chrysostomou}, {Curtis}, {Davis}, {Etxaluze}, {Francesco}, {Friberg},
  {Friesen}, {Greaves}, {Hatchell}, {Johnstone}, {Matthews}, {Matthews},
  {Matzner}, {Nutter}, {Rawlings}, {Roberts}, {Sadavoy}, {Simpson}, {Tothill},
  {Tsamis}, {Viti}, {Ward-Thompson}, {White}, {Wouterloot}, \&
  {Yates}}]{2010MNRAS...Graves}
{Graves}, S.~F., {Richer}, J.~S., {Buckle}, J.~V., {et~al.} 2010, \mnras, 409,
  1412

\bibitem[{{Hartmann} \& {Burkert}(2007)}]{2007ApJ...654..988H}
{Hartmann}, L. \& {Burkert}, A. 2007, \apj, 654, 988

\bibitem[{{Harvey} {et~al.}(2006){Harvey}, {Chapman}, {Lai}, {Evans}, {Allen},
  {J{\o}rgensen}, {Mundy}, {Huard}, {Porras}, {Cieza}, {Myers}, {Mer{\'{\i}}n},
  {van Dishoeck}, {Young}, {Spiesman}, {Blake}, {Koerner}, {Padgett},
  {Sargent}, \& {Stapelfeldt}}]{2006ApJ...644..307H}
{Harvey}, P.~M., {Chapman}, N., {Lai}, S.-P., {et~al.} 2006, \apj, 644, 307

\bibitem[{{Harvey} {et~al.}(2007){Harvey}, {Rebull}, {Brooke}, {Spiesman},
  {Chapman}, {Huard}, {Evans}, {Cieza}, {Lai}, {Allen}, {Mundy}, {Padgett},
  {Sargent}, {Stapelfeldt}, {Myers}, {van Dishoeck}, {Blake}, \&
  {Koerner}}]{2007ApJ...663.1139H}
{Harvey}, P.~M., {Rebull}, L.~M., {Brooke}, T., {et~al.} 2007, \apj, 663, 1139

\bibitem[{{Heitsch} {et~al.}(2009){Heitsch}, {Ballesteros-Paredes}, \&
  {Hartmann}}]{2009ApJ...704.1735H}
{Heitsch}, F., {Ballesteros-Paredes}, J., \& {Hartmann}, L. 2009, \apj, 704,
  1735

\bibitem[{{Heitsch} {et~al.}(2008){Heitsch}, {Hartmann}, {Slyz}, {Devriendt},
  \& {Burkert}}]{2008ApJ...674..316H}
{Heitsch}, F., {Hartmann}, L.~W., {Slyz}, A.~D., {Devriendt}, J.~E.~G., \&
  {Burkert}, A. 2008, \apj, 674, 316

\bibitem[{{Higuchi} {et~al.}(2010){Higuchi}, {Kurono}, {Saito}, \&
  {Kawabe}}]{2010ApJ...719.1813H}
{Higuchi}, A.~E., {Kurono}, Y., {Saito}, M., \& {Kawabe}, R. 2010, \apj, 719,
  1813

\bibitem[{{Kaas} {et~al.}(2004){Kaas}, {Olofsson}, {Bontemps}, {Andr{\'e}},
  {Nordh}, {Huldtgren}, {Prusti}, {Persi}, {Delgado}, {Motte}, {Abergel},
  {Boulanger}, {Burgdorf}, {Casali}, {Cesarsky}, {Davies}, {Falgarone},
  {Montmerle}, {Perault}, {Puget}, \& {Sibille}}]{2004A&A...421..623K}
{Kaas}, A.~A., {Olofsson}, G., {Bontemps}, S., {et~al.} 2004, Astronomy and
  Astrophysics, 421, 623

\bibitem[{{Kitsionas} \& {Whitworth}(2007)}]{2007MNRAS.378..507K}
{Kitsionas}, S. \& {Whitworth}, A.~P. 2007, \mnras, 378, 507

\bibitem[{{Klessen} {et~al.}(2005){Klessen}, {Ballesteros-Paredes},
  {V{\'a}zquez-Semadeni}, \& {Dur{\'a}n-Rojas}}]{2005ApJ...620..786K}
{Klessen}, R.~S., {Ballesteros-Paredes}, J., {V{\'a}zquez-Semadeni}, E., \&
  {Dur{\'a}n-Rojas}, C. 2005, \apj, 620, 786

\bibitem[{{Koda} {et~al.}(2006){Koda}, {Sawada}, {Hasegawa}, \&
  {Scoville}}]{2006ApJ...638..191K}
{Koda}, J., {Sawada}, T., {Hasegawa}, T., \& {Scoville}, N.~Z. 2006, \apj, 638,
  191

\bibitem[{{Larson}(1981)}]{1981MNRAS.194..809L}
{Larson}, R.~B. 1981, \mnras, 194, 809

\bibitem[{{Molinari} {et~al.}(2010){Molinari}, {Swinyard}, {Bally}, {Barlow},
  {Bernard}, {Martin}, {Moore}, {Noriega-Crespo}, {Plume}, {Testi}, {Zavagno},
  {Abergel}, {Ali}, {Anderson}, {Andr{\'e}}, {Baluteau}, {Battersby},
  {Beltr{\'a}n}, {Benedettini}, {Billot}, {Blommaert}, {Bontemps}, {Boulanger},
  {Brand}, {Brunt}, {Burton}, {Calzoletti}, {Carey}, {Caselli}, {Cesaroni},
  {Cernicharo}, {Chakrabarti}, {Chrysostomou}, {Cohen}, {Compiegne}, {de
  Bernardis}, {de Gasperis}, {di Giorgio}, {Elia}, {Faustini}, {Flagey},
  {Fukui}, {Fuller}, {Ganga}, {Garcia-Lario}, {Glenn}, {Goldsmith}, {Griffin},
  {Hoare}, {Huang}, {Ikhenaode}, {Joblin}, {Joncas}, {Juvela}, {Kirk},
  {Lagache}, {Li}, {Lim}, {Lord}, {Marengo}, {Marshall}, {Masi}, {Massi},
  {Matsuura}, {Minier}, {Miville-Desch{\^e}nes}, {Montier}, {Morgan}, {Motte},
  {Mottram}, {M{\"u}ller}, {Natoli}, {Neves}, {Olmi}, {Paladini}, {Paradis},
  {Parsons}, {Peretto}, {Pestalozzi}, {Pezzuto}, {Piacentini}, {Piazzo},
  {Polychroni}, {Pomar{\`e}s}, {Popescu}, {Reach}, {Ristorcelli}, {Robitaille},
  {Robitaille}, {Rod{\'o}n}, {Roy}, {Royer}, {Russeil}, {Saraceno}, {Sauvage},
  {Schilke}, {Schisano}, {Schneider}, {Schuller}, {Schulz}, {Sibthorpe},
  {Smith}, {Smith}, {Spinoglio}, {Stamatellos}, {Strafella}, {Stringfellow},
  {Sturm}, {Taylor}, {Thompson}, {Traficante}, {Tuffs}, {Umana}, {Valenziano},
  {Vavrek}, {Veneziani}, {Viti}, {Waelkens}, {Ward-Thompson}, {White},
  {Wilcock}, {Wyrowski}, {Yorke}, \& {Zhang}}]{2010A&A...518L.100M}
{Molinari}, S., {Swinyard}, B., {Bally}, J., {et~al.} 2010, \aap, 518, L100+

\bibitem[{{Myers}(2009)}]{2009ApJ...700.1609M}
{Myers}, P.~C. 2009, \apj, 700, 1609

\bibitem[{{Olmi} \& {Testi}(2002)}]{2002A&A...392.1053O}
{Olmi}, L. \& {Testi}, L. 2002, \aap, 392, 1053

\bibitem[{{Oort}(1954)}]{1954BAN....12..177O}
{Oort}, J.~H. 1954, \bain, 12, 177

\bibitem[{{Palla} \& {Stahler}(2002)}]{2002ApJ...581.1194P}
{Palla}, F. \& {Stahler}, S.~W. 2002, \apj, 581, 1194

\bibitem[{{Peretto} {et~al.}(2007){Peretto}, {Hennebelle}, \&
  {Andr{\'e}}}]{2007A&A...464..983P}
{Peretto}, N., {Hennebelle}, P., \& {Andr{\'e}}, P. 2007, \aap, 464, 983

\bibitem[{{Price} \& {Bate}(2007)}]{2007MNRAS.377...77P}
{Price}, D.~J. \& {Bate}, M.~R. 2007, \mnras, 377, 77

\bibitem[{{Price} \& {Monaghan}(2005)}]{2005MNRAS.364..384P}
{Price}, D.~J. \& {Monaghan}, J.~J. 2005, \mnras, 364, 384

\bibitem[{{Schneider} {et~al.}(2010){Schneider}, {Csengeri}, {Bontemps},
  {Motte}, {Simon}, {Hennebelle}, {Federrath}, \&
  {Klessen}}]{2010A&A...520A..49S}
{Schneider}, N., {Csengeri}, T., {Bontemps}, S., {et~al.} 2010, \aap, 520, A49+

\bibitem[{{Scoville} {et~al.}(1986){Scoville}, {Sanders}, \&
  {Clemens}}]{1986ApJ...310L..77S}
{Scoville}, N.~Z., {Sanders}, D.~B., \& {Clemens}, D.~P. 1986, \apjl, 310, L77

\bibitem[{{Strai{\v z}ys} {et~al.}(1996){Strai{\v z}ys}, {{\v C}ernis}, \&
  {Barta{\v s}i{\= u}te}}]{1996BaltA...5..125S}
{Strai{\v z}ys}, V., {{\v C}ernis}, K., \& {Barta{\v s}i{\= u}te}, S. 1996,
  Baltic Astronomy, 5, 125

\bibitem[{{Sugitani} {et~al.}(2010){Sugitani}, {Nakamura}, {Tamura},
  {Watanabe}, {Kandori}, {Nishiyama}, {Kusakabe}, {Hashimoto}, {Nagata}, \&
  {Sato}}]{2010ApJ...716..299S}
{Sugitani}, K., {Nakamura}, F., {Tamura}, M., {et~al.} 2010, \apj, 716, 299

\bibitem[{{Tasker} \& {Tan}(2009)}]{2009ApJ...700..358T}
{Tasker}, E.~J. \& {Tan}, J.~C. 2009, \apj, 700, 358

\bibitem[{{Vallee}(1995)}]{1995AJ....110.2256V}
{Vallee}, J.~P. 1995, \aj, 110, 2256

\end{thebibliography}

\onecolumn
\newpage
\begin{appendix}

\section{Time snapshots of the models presented}
\label{snapshots}

\begin{figure}[!h]
\centering
\hfill
\includegraphics[width=\textwidth]{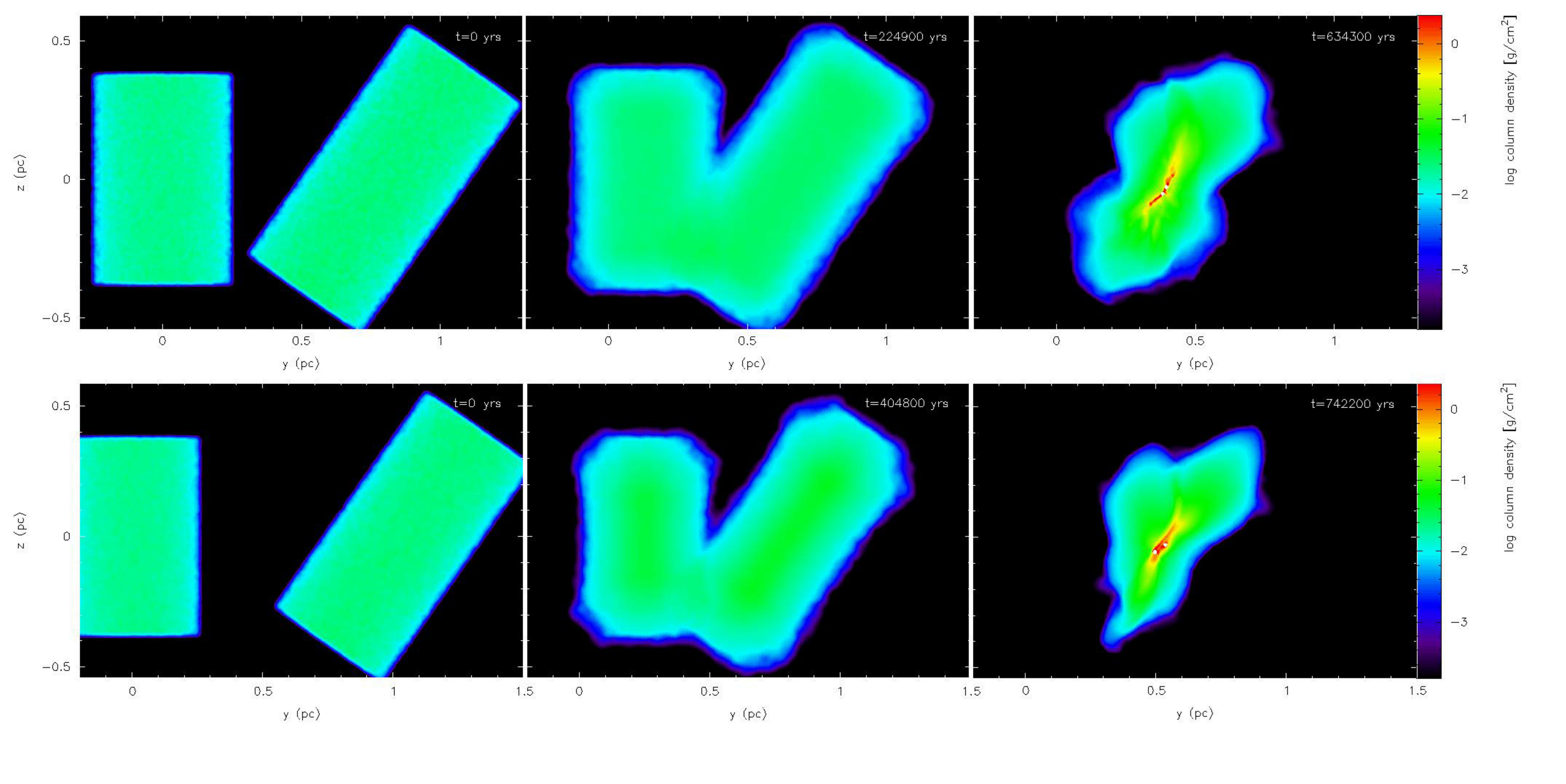}
\caption{\small{Three time snapshots of the total column density along the line of sight, for the non-turbulent runs: the centered collision, A$_{\mathrm{non-T}}$ (top) and off-centered collision, B$_{\mathrm{non-T}}$ (bottom). The three frames correspond to the beginning of the simulation (first frame), as soon as the cylinders start to collide (second frame) and when the first sink particle is formed (last frame).}}
\label{fig:nonturb_time}
\end{figure}

\begin{figure}[!ht]
\centering
\hfill
\includegraphics[width=\textwidth]{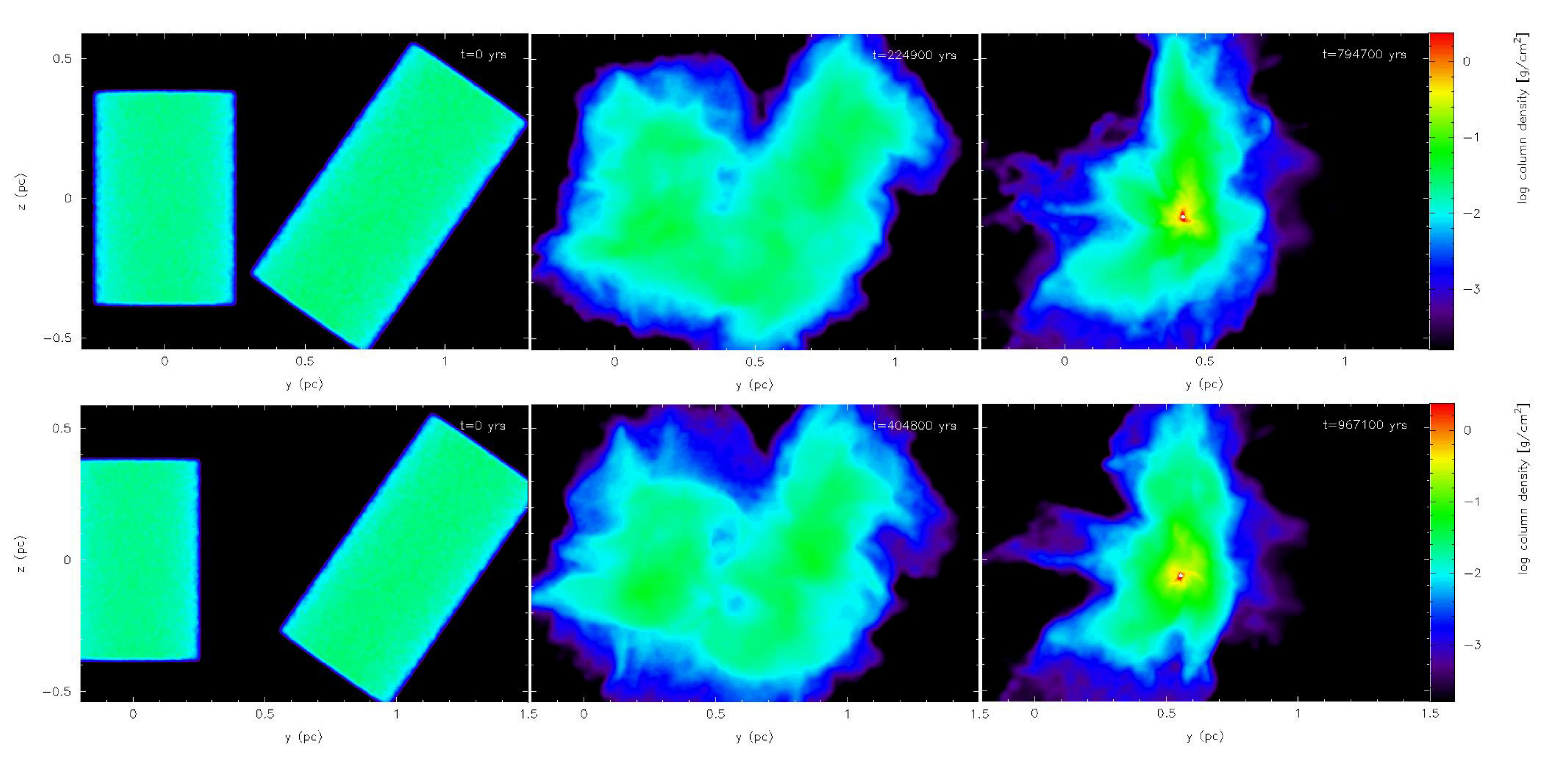}
\caption{\small{Three time snapshots of the total column density along the line of sight, for the turbulent runs: the centered collision, A$_{\mathrm{T}}$ (top) and off-centered collision, B$_{\mathrm{T}}$ (bottom). The three frames correspond to the beginning of the simulation (first frame), as soon as the cylinders start to collide (second frame) and when the first sink particle is formed (last frame).}}
\label{fig:turb_time}
\end{figure}

\begin{figure}[!ht]
\centering
\hfill
\includegraphics[width=0.98\textwidth]{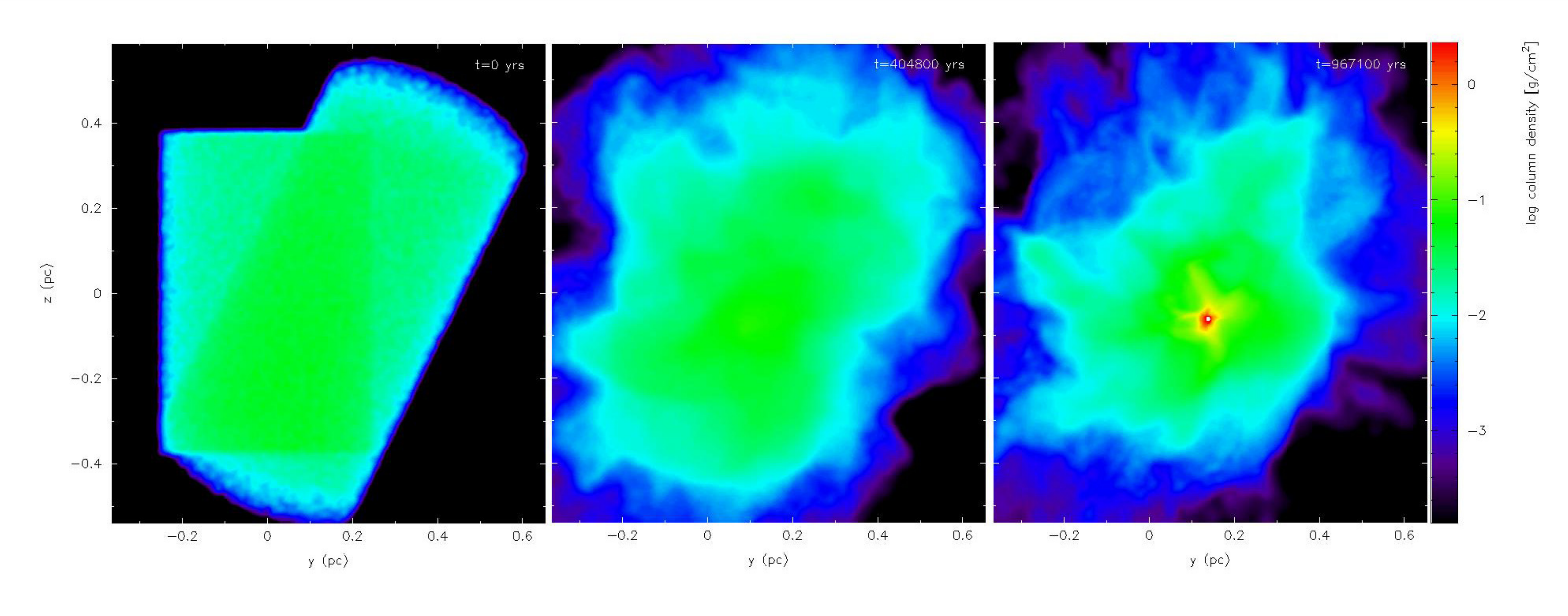}
\caption[Three time-steps of C$^_{\mathrm{T}}$ model]
	{\small{Three time snapshots of the total column density along the line of sight for C$_{\mathrm{T}}$. The three frames are as before. Note that this is the exact same model as B$_{\mathrm{T}}$, but viewed from a different perspective.}}
\label{fig:los_rot_time}
\end{figure}

\begin{figure}[!ht]
\centering
\hfill
\includegraphics[width=0.98\textwidth]{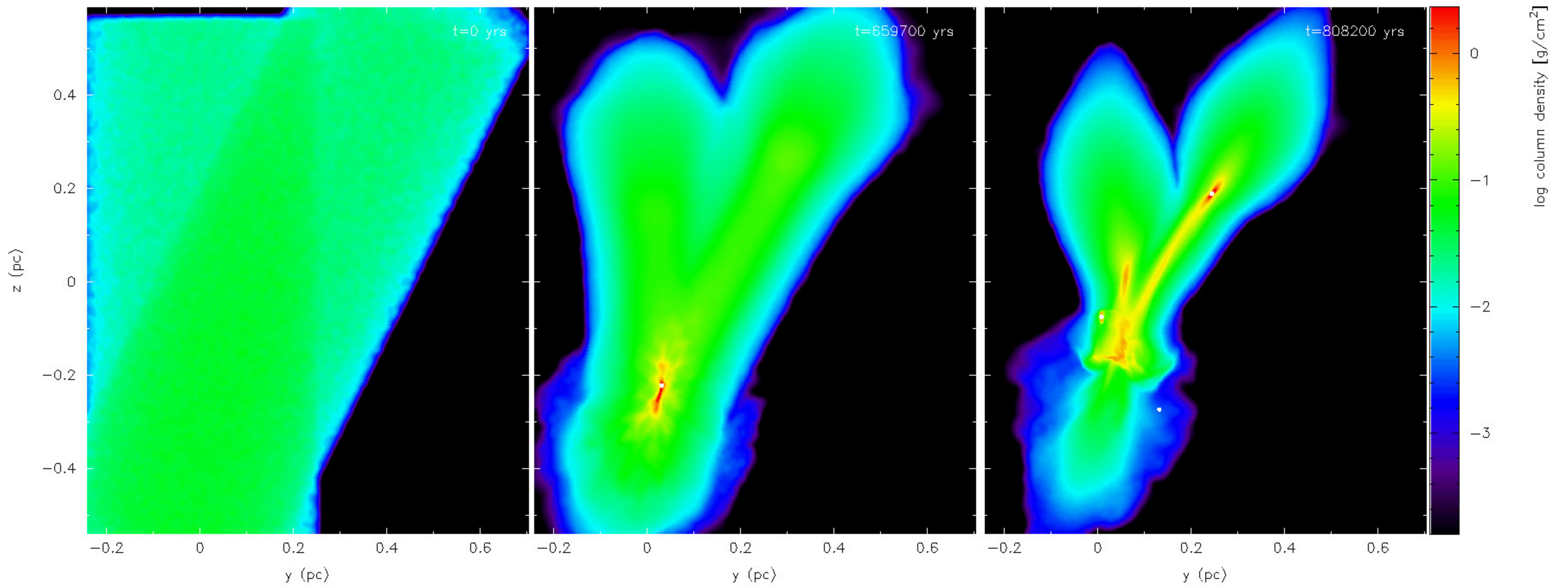}
\caption[Three time-steps of D$_{\mathrm{non-T}}$ model]
	{\small{Three time snapshots of the total column density along the line of sight for D$_{\mathrm{non-T}}$. The three frames are at the start of the simulation (left), when the first sink particle forms in the south (middle) and when a sink particle is formed in the north (right).}}
\label{fig:long_rot}
\end{figure}

Figures~\ref{fig:nonturb_time} to \ref{fig:long_rot} show three snapshots to illustrate the time evolution of the models presented in this paper. Each frame is a projected column density plot as seen from the chosen line of sight. The time (in years) is show in the top-right corner of each frame. Figure~\ref{fig:nonturb_time} shows three frames for the two non-turbulent runs presented, A$_{\mathrm{non-T}}$ and B$_{\mathrm{non-T}}$. Figure~\ref{fig:turb_time} shows the time evolution for the same models, A and B, but with turbulence. Figure~\ref{fig:los_rot_time} shows the snapshots of model C, which is essentially the same as model B$_{\mathrm{T}}$ only with a different line of sight. For these three runs, the last frame corresponds to the formation of the first sink particle and is the frame used to construct a datacube for comparison with the observations. Finally, Fig.~\ref{fig:long_rot} shows three snapshots for run D. Note that for this run we are showing frames after the formation of the first sink particle, to demonstrate where and when the second clump in the north is able to form a sink particle. In this case, this last frame was the one used for the comparison with the observations.

\end{appendix}

\end{document}